\documentclass[12pt,preprint]{aastex}

\def\geqsim{\lower.73ex\hbox{$\sim$}\llap{\raise.4ex\hbox{$>$}}$\,$}
\def\leqsim{\lower.73ex\hbox{$\sim$}\llap{\raise.4ex\hbox{$<$}}$\,$}

\newcommand{\myemail}{aross2@uiuc.edu}

\newcommand{\SDSSpixt}{{\em SDSSpix~}}
\newcommand{\SDSSpixb}{{\em SDSSpix}}

\slugcomment{{ApJ Accepted 04/19/2007}}
\shorttitle{N-Order Angular Galaxy Correlations in the SDSS }
\shortauthors{Ross et al.}

\begin{document}

\title{Higher-Order Angular Galaxy Correlations in the SDSS:  Redshift and Color Dependence of non-Linear Bias }

\author{Ashley J. Ross\altaffilmark{1}, Robert J. Brunner\altaffilmark{1,2},  Adam D. Myers\altaffilmark{1,2}}

\email{\myemail}

\altaffiltext{1}{Department of Astronomy,Ê 
University of Illinois at Urbana-Champaign,Ê 
Urbana, IL 61801}
\altaffiltext{2}{National Center for Supercomputing Applications,
Champaign, IL 61820}

\begin{abstract}
 We present estimates of the N-point galaxy, area-averaged, angular correlation functions $\bar{\omega}_{N}$($\theta$)  for $N$ = 2,...,7 for galaxies from the fifth data release of the Sloan Digital Sky Survey.  Our parent sample is selected from galaxies with $18 \leq r < 21$, and is the largest ever used to study higher-order correlations.  We subdivide this parent sample into two volume limited samples using photometric redshifts, and these two samples are further subdivided by magnitude, redshift, and color (producing early- and late-type galaxy samples) to determine the dependence of $\bar{\omega}_{N}$($\theta$) on luminosity, redshift, and galaxy-type.  We measure $\bar{\omega}_{N}$($\theta$) using oversampling techniques and use them to calculate the projected, $s_{N}$.  Using models derived from theoretical power-spectra and perturbation theory, we measure the bias parameters $b_1$ and $c_2$, finding that the large differences in both bias parameters ($b_1$ and $c_2$) between early- and late-type galaxies are robust against changes in redshift, luminosity, and $\sigma_8$, and that both terms are consistently smaller for late-type galaxies.  By directly comparing their higher-order correlation measurements, we find large differences in the clustering of late-type galaxies at redshifts lower than 0.3 and those at redshifts higher than 0.3, both at large scales ($c_2$ is larger by $\sim0.5$ at $z > 0.3$) and small scales (large amplitudes are measured at small scales only for $z > 0.3$, suggesting much more merger driven star formation at $z > 0.3$).  Finally, our measurements of  $c_2$ suggest both that $\sigma_8 < 0.8$ and $c_2$ is negative.
\end{abstract}

\keywords{Cosmology: observational; Galaxies: clustering, environment, star formation}

\section{Introduction}
An important subject in cosmology and galaxy formation is galaxy bias.  Galaxies trace the underlying dark matter that shapes the Universe, but there is no guarantee that they do so faithfully.  The galaxy bias defines the relationship between the clustering of dark matter and the clustering of galaxies.  Ideally, the relationship between the distribution of galaxies and dark matter is simple --- the ratio of the over-density of galaxies to the over-density of dark matter (i.e., the bias,  hereafter $b$, which is measured relative to clustering dark matter) is a constant factor independent of smoothing scale and the over-density of dark matter.  This ideal case is known as {\it{scale invariant linear bias}}.  If instead $b$ is a function of the overdensity, it is {\it{non-linear}}.  Significantly non-linear bias would suggest that galaxy formation is dependent on environment, as it would suggest that the efficiency of galaxy formation is dependent on halo mass (i.e., the dark matter over-density).

Assuming a non-linear bias, $b$ can formally be expanded into factors $b_N$ via a Taylor expansion, which reduces to linear bias when $b_1 = b$ and $b_{N} = 0$ for $N > 1$.  Several previoius studies have found that $b_2$ is nearly zero.  \cite{Pan05} found, by studying the monopole contribution to the 2dFGRS three-point correlation function, that, for galaxies with $-21\leq b_{j} \leq -20$, $b_{2} = -0.06_{-0.001}^{+0.003}$.  Likewise, \cite{H05} found, using the Sloan Digital Sky Survey (SDSS) spectroscopic galaxy catalog to calculate the bispectrum, that $c_{2}$ ($c_N = b_N/b_1$) is consistent with zero to within 10$\%$ for $\sigma_8$ = 0.9, however, they measure $c_2$ to be larger for smaller $\sigma_8$.  On the other hand, in an earlier study we (\citealt{Ross}, hereafter denoted Ross06) found $c_{2} = -0.30 \pm 0.10$ via determination of the $Nth$-order angular galaxy correlations for 11 million photometrically selected galaxies from the third data release (DR3) of the Sloan Digital Sky Survey (SDSS).  \cite{Gaz05} likewise found $c_{2} = -0.36_{-0.09}^{+0.13}$ for $b_{j} < 19.45$ galaxies in the 2dFGRS by calculating the redshift-space three-point correlation functions, a method that is independent of $\sigma_8$, the rms fluctuation in the matter density averaged in a sphere of radius 8 Mpc, .  

Determining $\sigma_8$ is necessary, because, as yet, there is no clear consensus.  When measured from only the third year WMAP (WMAP3) data, $\sigma_8 = 0.744^{+0.050}_{-0.060}$ \citep{Sper06}.  Further constraining WMAP3 data via inflationary models, $\sigma_8=0.700^{+0.063}_{-0.065}$.  When combined with the SDSS power-spectrum data, however, this increases to $0.772^{+0.036}_{-0.048}$, and when combined with the 2dFGRS power-spectrum data it decreases to $0.737^{+0.033}_{-0.045}$ \citep{Sper06}.  Furthermore, the SDSS power spectrum by itself determines $\sigma_8 = 0.842^{+0.069}_{-0.058}$ \citep{Teg04}.  Methods that employ clusters of galaxies typically measure smaller values of $\sigma_8$.  For example, \cite{Vod04} found $\sigma_8 = 0.72\pm0.04$ by using a cluster baryon mass function.  

Independent measurements of $b_1$ and $\sigma_8$ are not possible using correlation measurements, as $\sigma_8$ serves as the normalization to the model power spectra one must use to determine $b_1$.  The value of $b_2$, however, is not degenerate with $\sigma_8$, as it will change the {\it shape} of correlation measurements.  A novel technique to provide an independent measure of $\sigma_8$ is to capitalize on the independence of $b_2$.  By determining both $b_2$ and $c_2$, $\sigma_8$ can be constrained.  Determining the value of $c_2$ alone places loose constraints on the value of $\sigma_8$.

Our most dramatic findings in Ross06 were that significant non-linear bias was required to explain the observed clustering of early- and late-type galaxies.  The results determined that early-type galaxies' $c_2$ was larger than that of late-type galaxies by an absolute difference of 1.0$\pm$0.13.  \cite{Cr06} have since confirmed this measurement --- that red galaxies have a higher $c_2$, via their determination of higher-order correlations in the 2dFGRS, though they found a more modest difference of 0.36$\pm$0.17.  In another study, \cite{Nish06} found a difference of closer to 0.5 when studying the bispectrum of galaxies selected from the SDSS data release 4 (DR4) spectroscopic survey.  It is not surprising that differences are observed between these three studies, as each uses a different color cut to separate the red and blue samples.  The existence of non-linear bias between galaxy types was first found by \cite{Gaz92}, when it was discovered that the higher-order clustering of galaxies selected by optical surveys (CfA and SSRS) differs from the clustering of galaxies selected by the infrared survey IRAS (galaxy selection in IRAS was biased towards late-type galaxies).  This was confirmed by \cite{Fry93} and interpreted as IRAS (and thus late-type) galaxies having a negative $c_2$ relative to optically selected galaxies.  

Clearly, galaxy type affects the non-linear nature of the bias.  The linear bias is also strongly dependent on galaxy type, as red, early-type, galaxies have consistently been shown to cluster more strongly (e.g., \citealt{W98,N02,Ma03,Z05,Ross,Cr06}) than blue, late-type, galaxies.  Luminosity has been shown to scale proportionally to $b_1$ (e.g., \citealt{Ma03,Z05}), which logically follows if one expects more luminous objects to generally be more massive.  It is also likely that the properties of galaxies, and thus $b$, evolve over cosmic time.  Therefore, a true characterization of galaxy bias, both linear and non-linear, requires quantifying any dependence on galaxy type, luminosity, and redshift.

As a result, we employ the wealth of information available from SDSS DR5 to follow-up and vastly improve our DR3 measurements on the nature of galaxy bias.  Apart from Ross06, four other studies have used SDSS photometric data to calculate higher-order correlations:   \cite{GazEDRa},   \cite{GazEDR}, and \cite{Sza02} measured higher-order correlations for galaxies in the SDSS Early Data Release (EDR) and found that measurements using SDSS were consistent with previous results and were free of any systematics.  \cite{Bl06} measured the DR1 higher-order correlations and found good agreement with simulation.  The Ross06 measurements employed the largest data set ever used to calculate higher-order correlations with over 11 million galaxies from DR3.  The SDSS DR5 offers an opportunity to significantly improve the DR3 results, as it offers $\sim70\%$ more data and accurate photometric redshift catalogs complete with rest-frame absolute-magnitudes.  This wealth of data enables a quantification of the dependence of the $Nth$-order correlations of galaxies on luminosity, type, and redshift.  

In this paper, we, therefore, calculate and analyze the area-averaged angular $N$-point correlation functions using SDSS DR5 galaxies, up to seventh-order.  Our methodology is explained in \S 3.  This present work offers a significant improvement over Ross06, due to the fact that we use $\sim70\%$ more data, extend the range of measurement by an order of magnitude, make measurements using volume limited samples, and employ improved theoretical modeling.  Our main sample of galaxies is used for comparison with the Ross06 measurements, and is split into the same five subsamples as in Ross06,  a process which is described in \S 2.  The measurements made using DR5 are presented in \S 4, where we illustrate the superiority of DR5 over DR3.  We use the SDSS DR5 {\tt PhotoZ} table to create two volume limited samples, allowing us to investigate the evolution and luminosity dependence of $Nth$-order correlations in \S 5.  Model, area-averaged correlation functions are calculated by integrating over theoretical power spectra and an appropriate redshift distribution, allowing for the calculation of first- and second-order bias parameters.  In \S 6, we present our measurements of the higher-order (non-linear) bias and determine their dependence on galaxy-type, luminosity, and redshift.  

We adopt the cosmology ($\Omega_m$, $\Omega_{\Lambda}$, $h$, $\Gamma$) $=$ (0.28, 0.72, 0.7, 0.15), where $\Gamma$ is the shape parameter, based on recent supernovae, large-scale structure, and CMB measurements (e.g., \citealt{Riess04, Sper06, Cole05}).  This value of the shape parameter arises naturally (see, e.g., Equations 30 and 31 of \cite{Eis98}) for a baryon fraction of $\Omega_b/\Omega_m = 0.185$ (e.g., \citealt{Cole05}).

\section{Data}
The data analyzed herein were taken from the SDSS DR5 \citep{Ab05}.  This survey obtains wide-field CCD photometry \citep{C} in five passbands ($u,g,r,i,z$; e.g., \citealt{F}).  The entire DR5 represents 8000 square degrees of observing area.  We selected galaxies with positions lying in the Northern, contiguous portion of the SDSS from the DR5 {\tt PhotoPrimary} database, and further constrained the sample (using the \citealt{Sc} dust maps) to have reddening-corrected magnitudes in the range $18 \leq r <  21$.  Further, significant masking was required to account for bright stars and areas of high reddening and poor seeing (see \S \ref{sec:masks}).  This produced a set of over 18 million galaxies (18,532,911) at a median redshift of about 0.31.  This is by far the most galaxies used to conduct this type of measurement, and represents over a 70$\%$ increase in the number of objects used in our previous DR3 measurements.  We split this sample of galaxies into the same five sub-samples as in Ross06 (three samples constrained to the magnitude ranges $18 \leq r <  19$, $19 \leq r <  20$,  and $20 \leq r <  21$, and two color-selected samples defined by $u-r > 2.2$ for early-type and $u-r \leq 2.2$ for late-type galaxies).  These samples are used primarily for comparison with the DR3 measurements, however, due to our improved measurement techniques, they also sample substantially smaller and larger scales than Ross06.  

\subsection{Creation of Volume Limited Samples}
\begin{figure}
\plotone{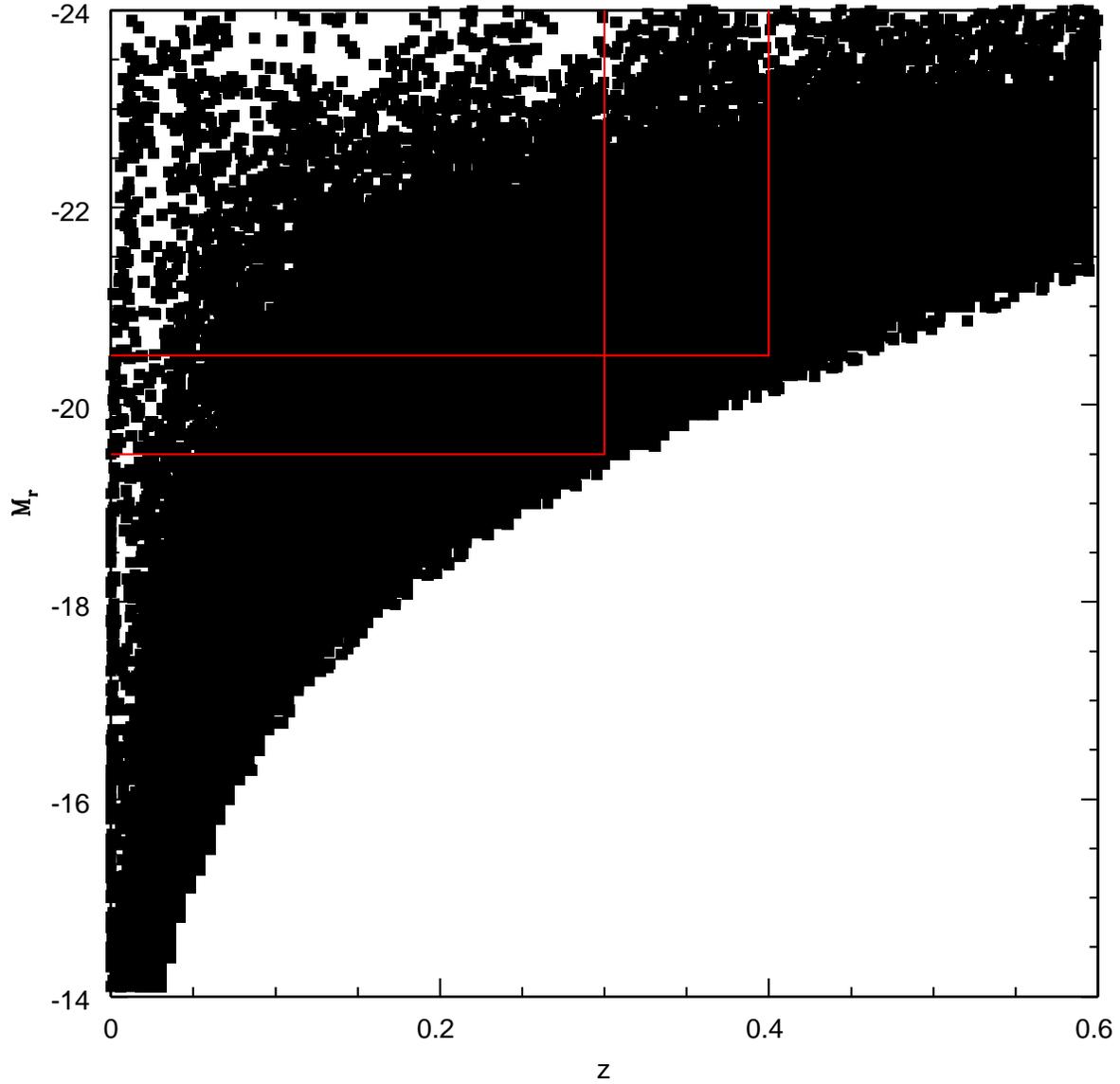}
\caption{Absolute $r$-band magnitude, $M_r$, versus redshift, $z$, for galaxies in the DR5 {\it{PhotoZ}} catalog.  The well-defined edge allows for the simple creation of volume limited samples.  The magnitude--redshift areas of the two volume limited samples analyzed herein, $z3$ and $z4$, are outlined in red. }
\label{fig:MvZ}
\end{figure}

Our primary analysis is divided between two volume limited samples created following the methods outlined in \cite{Bud03}.  Galaxies with $18 < r < 21$ are taken from the DR5 {\tt PhotoZ} table and matched to galaxies in the DR5 {\tt PhotoPrimary} table.  Using the rest-frame absolute $r$-band magnitudes, $M_r$, given for each galaxy in the {\tt PhotoZ} table, we display the $M_r$ -- $z$ plane in Figure \ref{fig:MvZ}.  As can be seen, there is a definite locus that defines the limiting absolute magnitude for a given redshift.  To create a volume limited catalog to a given redshift, $z_L$, we simply select galaxies with $z \leq z_L$ that are also intrinsically brighter than the limiting magnitude defined by the locus of points displayed in Figure \ref{fig:MvZ}.  We focus our analysis on two volume limited samples that have roughly similar numbers of galaxies.  One sample we take is limited to a redshift of 0.4, requiring that $M_r < -20.5$.  This sample, hereafter denoted as $z4$, contains nearly three and half million galaxies (3,380,553) after masking (see \S \ref{sec:masks}).  Our other volume limited sample, hereafter denoted $z3$, is limited to a redshift of 0.3, with $M_r < -19.5$, and contains nearly four million objects (3,980,652) after masking.

\section{Methodology}
\subsection{Angular Correlation functions}
We estimate N-point area-averaged angular correlation functions, $\bar{\omega}_{N}$($\theta$), using a counts-in-cells technique (e.g., Ross06).  This basically involves calculating the statistical moments of the over-densities contained in equal-area cells.   The over-density for cell $i$ is defined as
 \begin{equation}
 \delta_{i} = \frac{\bar{n} - n_{i}}{\bar{n}}
 \end{equation}
\noindent where $\bar{n}$ is the average number of galaxies in a cell and $n_{i}$ is the number of galaxies in cell $i$.  The remaining details and equations required to determine $\bar{\omega}_{N}$($\theta$) are found in Ross06.

In a hierarchical model (e.g., \citealt{Gro77,Sza92,Gaz94}), higher-order correlations can be expressed in terms of the two-point correlation function, and the volume-averaged correlations are given by
\begin{equation}
\bar{\xi}_{N}(R) = S_{N}[\bar{\xi}_{2}(R)]^{N-1}
\end{equation}
\noindent where $S_{N}$ is the hierarchical amplitude.  In a similar manner, we can define the analogous relationship for the area-averaged angular correlations
 \begin{equation}
s_{N} \equiv  \frac{\bar{\omega}_{N}(\theta)}{[\bar{\omega}_{2}(\theta)]^{N-1}}
\end{equation}

The hierarchical amplitudes of the higher-order moments encode much of the pertinent information on the distribution of the data. These amplitudes, therefore, embody the central analysis of this paper.

\subsection{Redshift Distributions}
\begin{figure}
\plotone{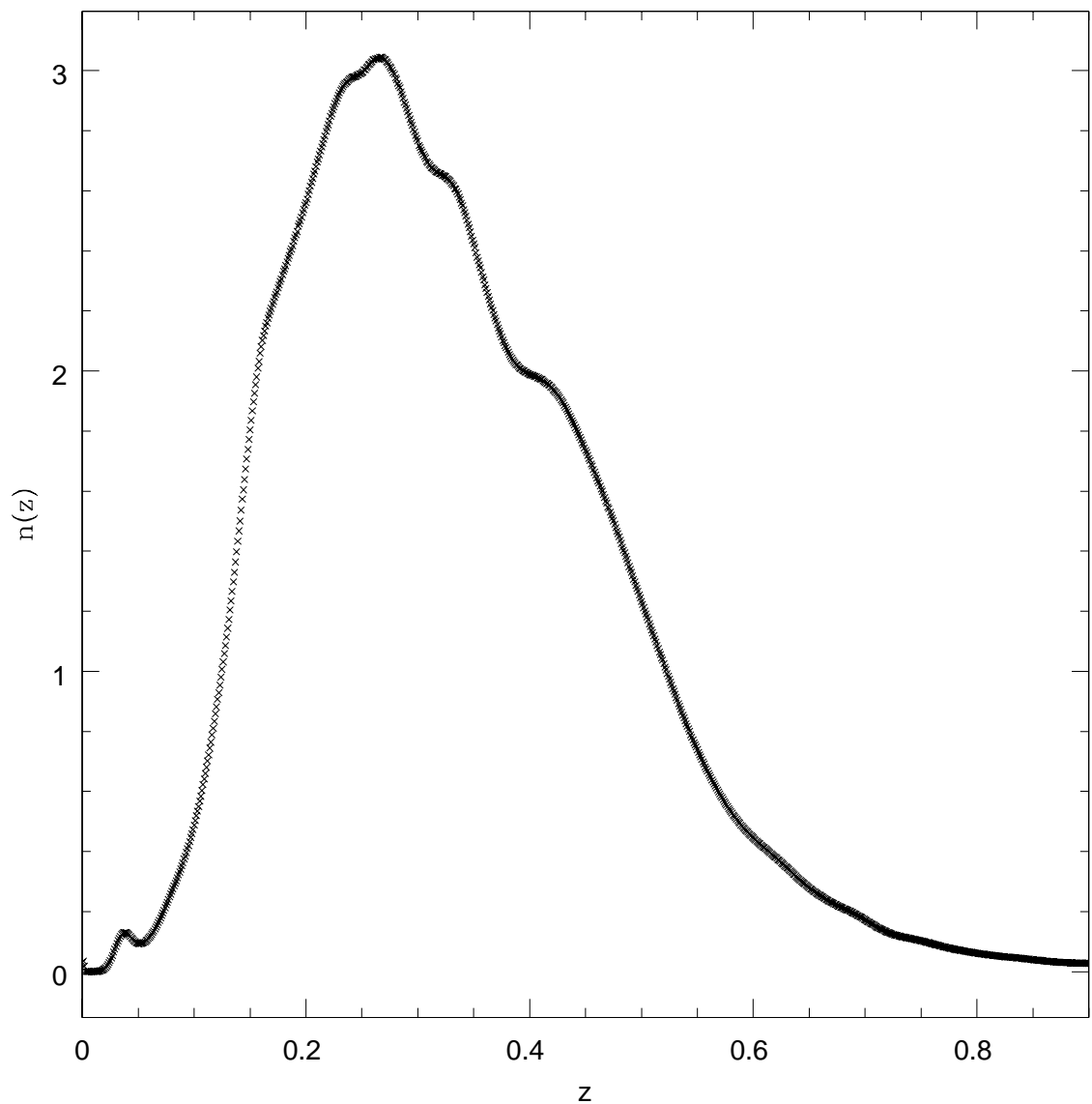}
\caption{Normalized number of galaxies, $n(z)$, in redshift bins $\Delta z = 10^{-3}$ determined by using the photometric redshifts of galaxies with $18 \leq r < 21$ in the DR5 {\tt{Photoz}} Table.  This curve defines the redshift selection function we use to transform between real-space theory and our projected measurements.  Similar curves are constructed for all other galaxy samples we analyzed herein.}
\label{fig:dndz}
\end{figure}

In order to compare angular measurements to theoretical models, it is necessary to determine $dn/dz$, which we accomplish using the galaxies' photometric redshifts (see \S 2.1).  We construct $dn/dz$ by using each published redshift and its error (rejecting any with error greater than twenty percent) to create a probability density function (PDF).  The PDFs for each redshift are combined to produce the expected number of objects ($n$) in a redshift bin of width 0.001.  This distribution is then normalized and interpolated over in order to estimate $\frac{dn}{dz}$.  For SDSS galaxies with $18 \leq r < 21$, the resulting normalized $n(z)$ is plotted in Figure \ref{fig:dndz}.  The distribution of $n(z)$ is smooth and roughly Gaussian. For all subsamples (volume limited and otherwise, see \S 2), we likewise use the corresponding DR5 photometric redshifts to obtain estimates of their individual redshift selection functions.

\subsection{Pixelization}
\label{sec:pixel}

The pixelization schemes we employ are nearly identical to those used in Ross06.  Basically, we reimplemented the \SDSSpixt pixelization scheme originally developed by Tegmark, Xu, and Scranton\footnote{http://lahmu.phyast.pitt.edu/$\sim$scranton/SDSSPix/}, as described in detail in Ross06.  Increased computing resources have allowed efficient usage of \SDSSpixt at smaller scales, allowing us to make measurements for $\theta > 0.02^{0}$. The {\it{striped}} method (see Ross06) is again employed, but only for scales between 0.005 and 0.04 degrees (we note that we are able to probe smaller scales than in Ross06).

For larger angular scales an {\it oversampling} technique (e.g., \citealt{Sza02}) is applied such that every angular scale uses the same number of data cells.  Using the base resolution pixels produced by \SDSSpixb, we can make cells equivalent to any angular scale $n\theta_b$ at every single base pixel (where $n$ is any integer greater than or equal to 2 and $\theta_b$ is the angular scale at the base resolution).  Thus, at a large scale the cells are highly overlapping, allowing more information to be extracted, which allows for more precise calculations at large scales.  We therefore perform calculations for $\theta < 20^{0}$.  We have verified that the results using the striped method are consistent with using this {\it oversampling} implementation at scales between 0.02 and 0.1 degrees.  From here on, we will refer to the method used for small angular scales as the {\it{striped}} method and to the method for large angular scales as the {\it oversampling} method.  Oversampling increases the covariance, but this is not a problem since we perform a full covariance analysis for all parameters we attempt to measure.

\subsection{Errors and Covariance}
\label{sec:err}

We compute errors and covariance matrices using a jackknife method (e.g., \citealt{Scr02}), with inverse-variance weighting for both errors (e.g., \citealt{Mye05,Mye06}) and covariance (e.g., \citealt{Mye07}), nearly identical to the method described in detail in Ross06.  The jackknife method works by creating many subsamples of the entire data set, each with a small part of the total area removed.  For the {\it{striped}} method, we utilize the natural geometry of the SDSS.  Each of the 29 different stripes in the DR5 forms a natural subset of the overall data.  The covariance matrix is calculated using each of the possible subsamples of DR5 that is made up of 28 stripes.  For the angular scales that are calculated by the {\it{striped}} method, we find that these 29 subsamples are sufficient to create a stable covariance matrix.  For the larger angular scales probed by the {\it oversampling} method, we find that 20 jack-knife subsamplings are sufficient to create a stable covariance matrix.  These 20 subsamples are created by simply eliminating a contiguous grouping of 1/20th of the unmasked pixels in 20 separate areas.  To properly constrain fit parameters, we minimize the $\chi^2$ using our covariance matrixes via
\begin{equation}
\chi^2 = \sum_{i,j}[\bar{\omega}(\theta_i) - \bar{\omega}_{m}(\theta_i)]C_{i,j}^{-1}[\bar{\omega}(\theta_j) - \bar{\omega}_{m}(\theta_j)]
\label{eq:Chi}
\end{equation}
where $C$ is the covariance matrix and $i$ and $j$ refer to the $i^{th}$ and $j^{th}$ jackknife subsample.

\subsection{Masks}
\label{sec:masks}
We will generally refer to useful observational information (such as seeing and Galactic extinction values) across each pixel in our schema (see \S\ref{sec:pixel}) as forming a mask of that information. The DR5 area required significant masking which we performed in the same manner as in Ross06.  Pixels at the base resolution are discarded if they intersect the standard SDSS imaging mask, have a mean reddening $A_{r} > 0.2$, have a mean seeing greater than 1.$\arcsec$5, or intersect the Ross06 mask for galaxy M101.  Cells at scales above the base resolution have their over-densities corrected for the fractional area of the pixel in the same manner as in Ross06, thus  
\begin{equation}
 \delta_{i} = \frac{\bar{n} - \frac{n_{i}}{\Delta_{i}}}{\bar{n}}
 \end{equation}
\noindent where $\Delta_{i}$ is the fractional area of cell $i$.  As before, we did not find any systematic variation in correlation measurements as a function of stripe.

\section{Area-averaged Correlation Functions and Hierarchical Amplitudes: Complete Sample}

\begin{figure}
\plotone{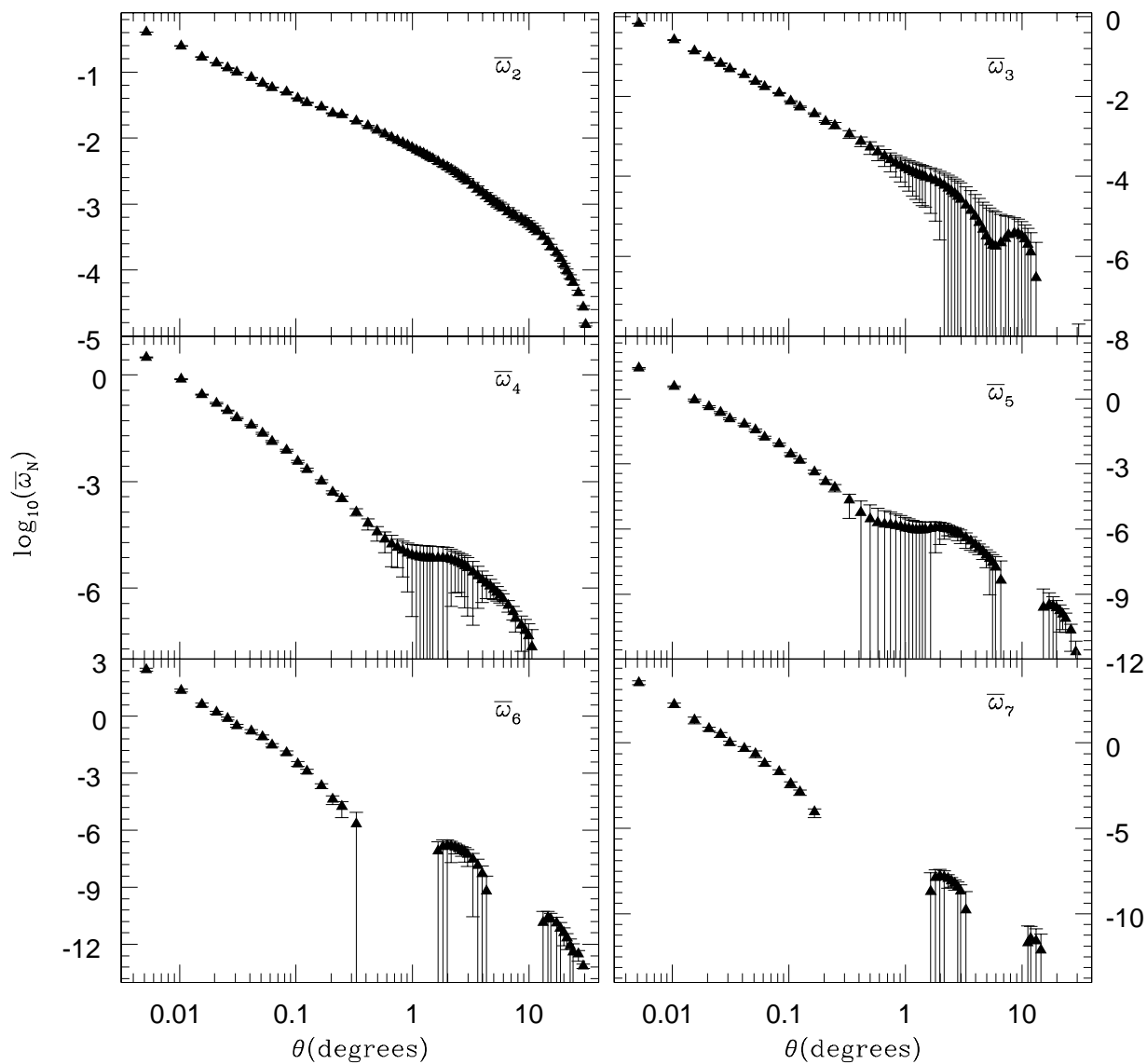}
\caption{The area-averaged, angular, N-point correlation functions for $18 \leq r < 21$  galaxies for $N$ = 2,.....,7 (left to right, top to bottom).  Note the logarithmic scaling.  Approximate power-law behavior is observed for all $N$ to about 0.3 degrees, above which, however a power law clearly is no longer valid.}
\label{fig:corr}
\end{figure}

Figure \ref{fig:corr} shows the area-averaged correlation functions for $N \leq 7$ determined for galaxies with $18 \leq r < 21$.  Errors for each point were determined by the jackknife method (see Ross06).  Compared to the Ross06 results, we now sample at more angular scales, and the maximum and minimum angles are extended, resulting in an extra order of magnitude in angular coverage.  For each $N$, the correlation function has a shape roughly consistent with a power-law to about 0.3$^{o}$.  At larger scales, there exist obvious features that are not adequately represented by a power law.  Looking at $\bar{\omega}_{2}$, a power-law representation appears as if it may be appropriate to 1$^{o}$.  Assuming a form $A\theta^{1-\gamma}$, the $\chi^{2}$ best fit for $N = 2$ over the angular range $0.02<\theta < 1^o$ is $A =  (7.2 \pm 0.1)\times10^{-3}$ and $\gamma = 1.770 \pm 0.005$.  These results are consistent with previous measurements (see, e.g., \citealt{Gaz94,Con02, FR05}, Ross06).  The $\chi^{2}$ value is 433.6, however, meaning that a power-law form is inappropriate over these scales.  In order to produce fits that are not rejected, fits must be performed over ranges that are less than half a dex.  We caution, therefore that our power-law measurement is useful only for comparison purposes.

\begin{figure}[hbtp]
\plotone{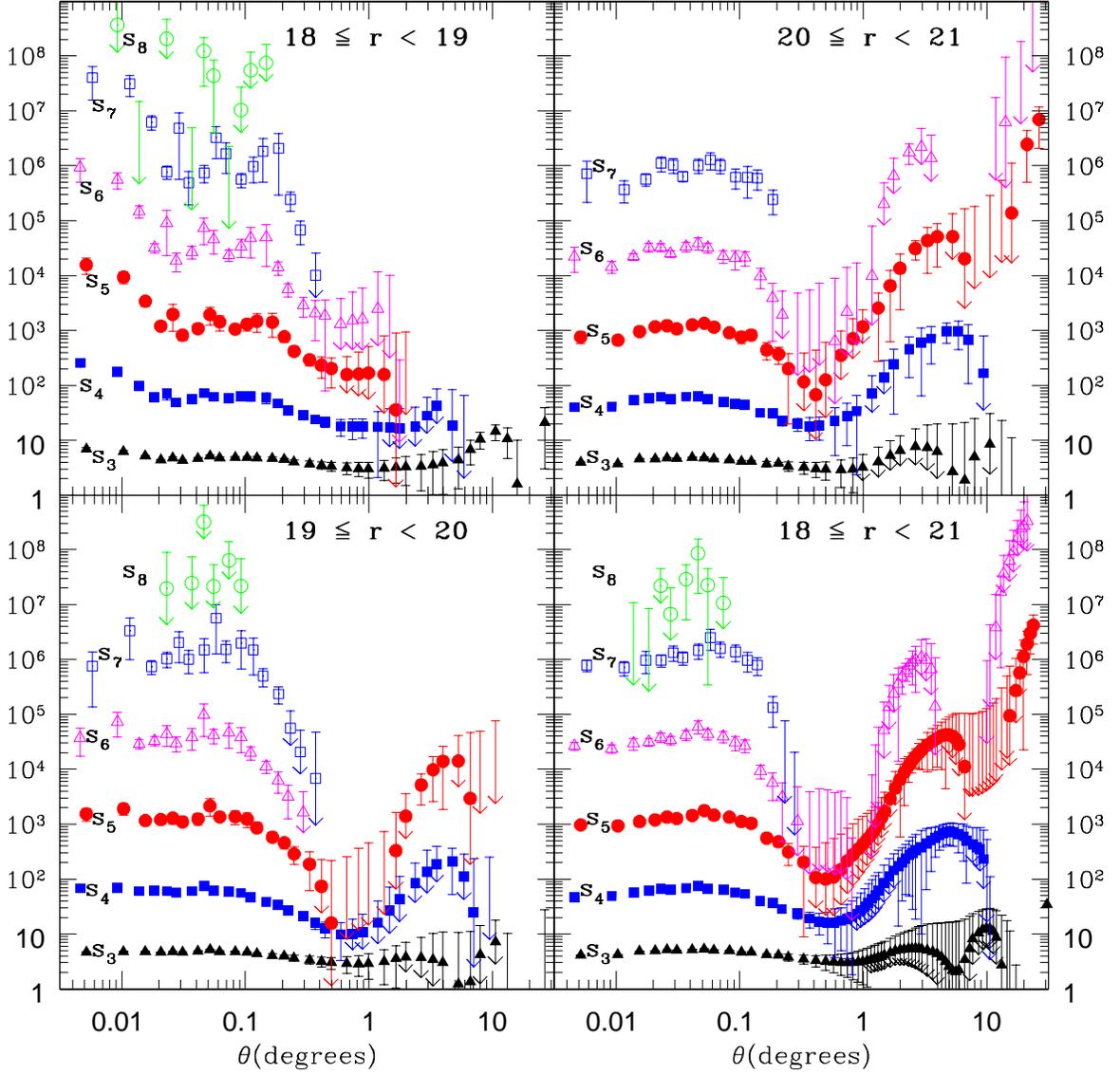}
\caption{The hierarchical amplitudes $s_{3}$ through $s_{7}$ for galaxies with $18 \leq r < 19$ (top left), $19 \leq r < 20$ (bottom left), $20 \leq r < 21$ (top right), and $18 \leq r < 21$ (bottom right). The $\theta$ values for each amplitude have been shifted and data with extremely large errors are omitted for clarity.  The amplitudes are roughly constant, consistent with the hierarchical model, however significant structure is present.}
\label{fig:9}
\end{figure}

The hierarchical amplitudes $s_{N}(\theta)$ for $N \leq 7$ measured in magnitude ranges $18 \leq r < 19$,  $19 \leq r < 20$, $20 \leq r<21$, and $18 \leq r < 21$ are shown in Figure~\ref{fig:9}.  Compared to the results of Ross06, the measurements made at scales greater than one degree are far more significant, but overall the measurements are quite consistent.  We also probe smaller scales, showing that at scales less than 0.02$^0$ the $s_N$ measured for $18 \leq r < 19$ increase as the angular scales decreases.   

As in Ross06, we employ the simple color criteria determined by \cite{Stra01} to separate early-type ($u-r > 2.2$) and late-type ($u-r \leq 2.2$) galaxies from SDSS photometric data.  Using this color cut and the magnitude restriction $18 \leq r < 21$, we separate our sample into early- and late-type galaxies and repeat our N-point measurements for these two samples.  We find that DR5 contains about 25\% more late-type galaxies (10,284,575) than early-type galaxies (8,248,336), which is nearly the same proportions as in Ross06.

\begin{figure}[hbtp]
\plotone{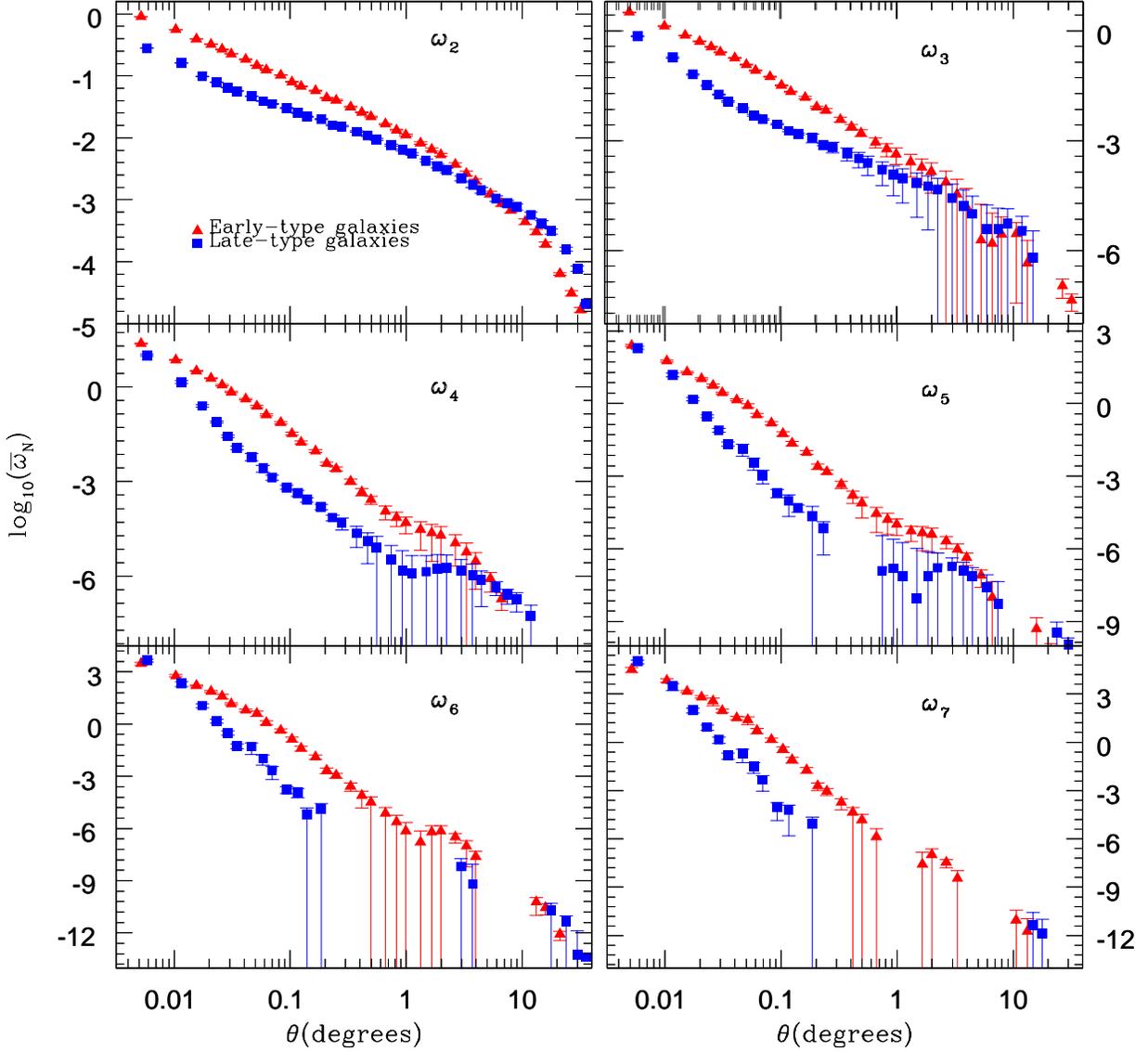}
\caption{The area-averaged angular N-point correlation functions for $18 \leq r < 21$ early-type (red, triangles) and late-type (blue, squares) galaxies (types determined by the \cite{Stra01} $u - r$ color cut), for $N$ = 2,.....,7 (left to right, top to bottom).  Note the logarithmic scaling.  Clear differences between the two galaxy types are evident in both the slopes and amplitudes of these measurements.}
\label{fig:12}
\end{figure}

\begin{figure}[hbtp]
\plotone{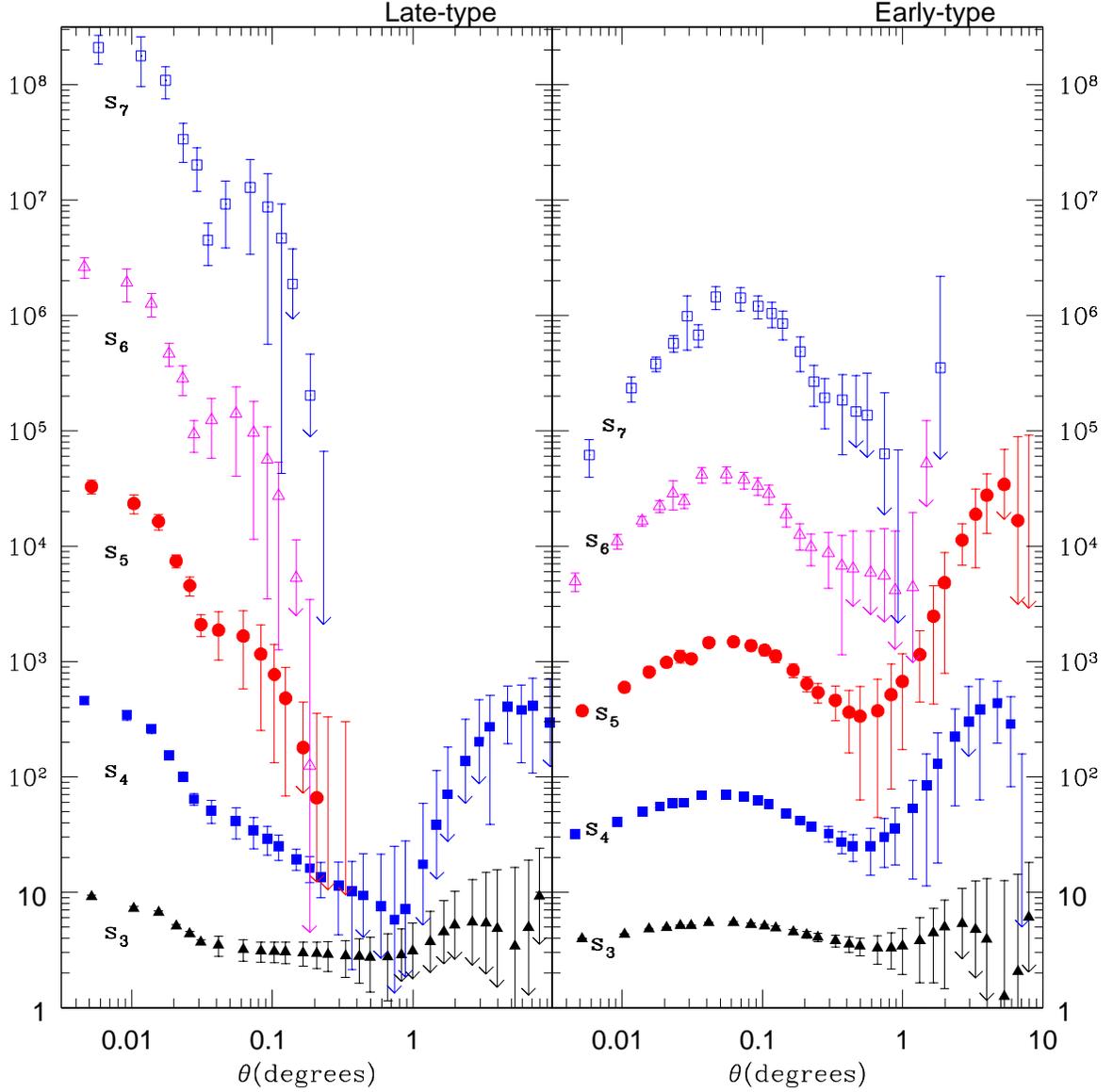}
\caption{The hierarchical amplitudes $s_{3}$ through $s_{7}$ for late-type (left) and early-type (right) galaxies, with types determined by the \cite{Stra01} $u - r$ color cut. The $\theta$ values for $s_{4}$ and $s_{7}$ have been shifted and data with extremely large errors are omitted for clarity.  At large scales, the early-types have significantly larger amplitudes than the late-types, implying a significant difference in bias between the two samples.}
\label{fig:13}
\end{figure}

 Figures \ref{fig:12} and \ref{fig:13} show the results of the $\bar{\omega}_N$ and $s_N$ measurements for early- and late-type galaxies.  The early-type galaxies clearly show stronger clustering at scales $\theta < 5^{0}$, in agreement with previous results (e.g., \citealt{W98,Z02,N02,Ma03}), but at large scales ($\theta >$ 5$^0$), the amplitudes are nearly equal.  This suggests that their respective bias might be scale dependent (see \S \ref{sec:B1}).  In Ross06 we were not able to probe such large scales, otherwise our overall results remain consistent.  The correlation functions display roughly power-law behavior for all $N$, but there is significant structure that becomes more pronounced as $N$ grows.  The $s_N$ of the late-type galaxies show smaller amplitudes for $\theta > 0.1^{0}$, suggesting that higher-order bias terms are significant.  At the largest scales ($\theta >$ 2$^0$), $s_3$ and $s_4$ appear nearly identical for early- and late-types, yet in this regime, the errors begin to dominate the measurements.
 
As found in Ross06, the late-type correlation measurements show extremely interesting behavior.  At angular scales between 0.01$^0$ and 0.1$^0$ the late-type galaxies exhibit slopes that increase as the scale decreases.  Probing smaller scales, this relationship appears to turn over, as evidenced by the $s_N$ measured at $\theta <$ 0.02$^0$.  For $N =$ 6 and 7, there is a dramatic loss of signal at angular scales greater than 0.2$^0$.  A similar loss in signal is shared by the correlation measurements made by using all galaxies, but the measurements made by using early-type galaxies do not show this dramatic loss of signal.  This suggests that there is an intrinsic property of late-type galaxy clustering that is so strong it dominates the measurement for $N \geq$ 6.

\begin{figure}[hbtp]
\plotone{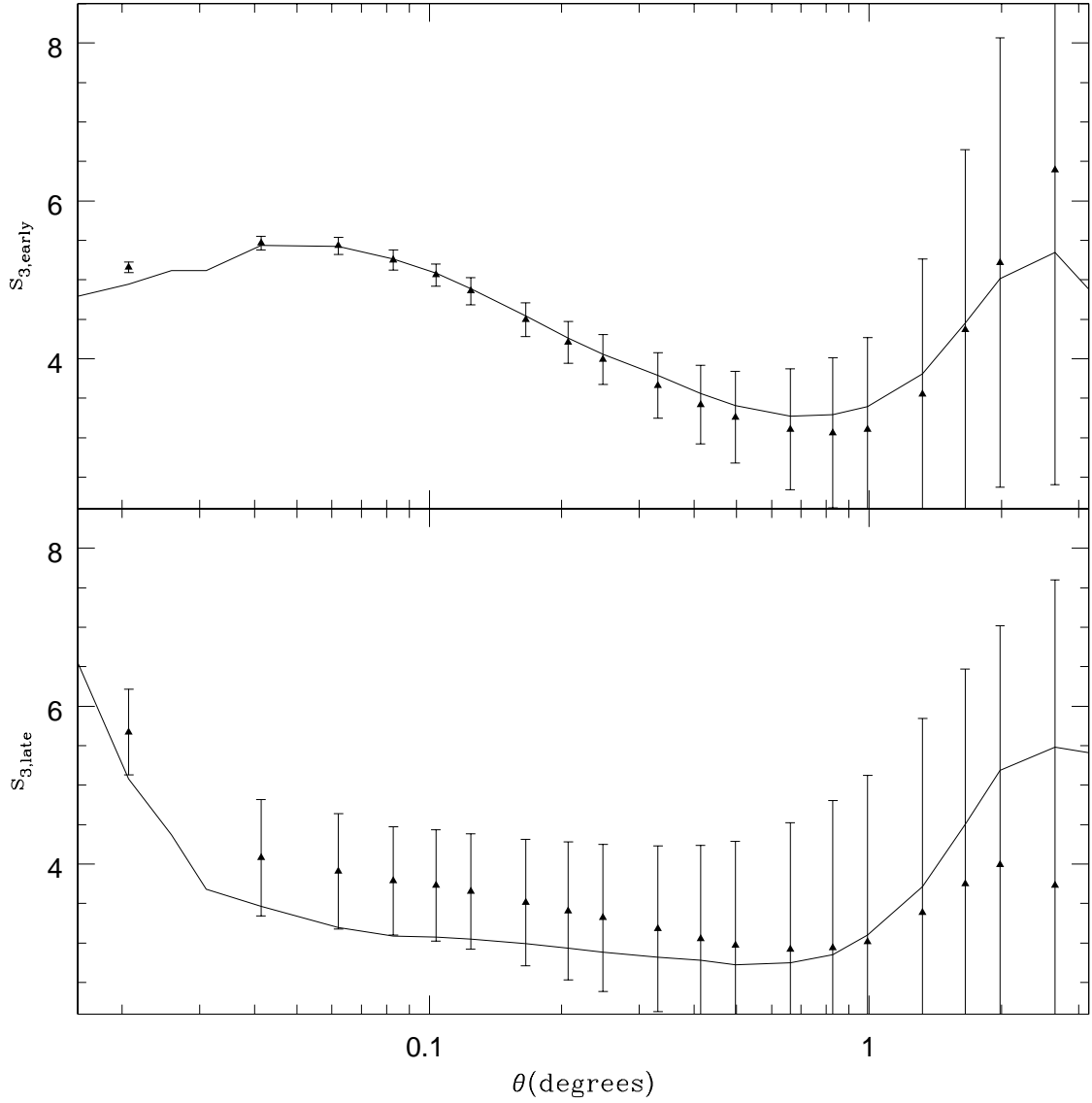}
\caption{The hierarchical amplitude $s_{3}$ for late-type (bottom) and early-type (top) galaxies. The solid line shows the measurement made by using galaxies with types determined by the \cite{Stra01} color cut, while the points display galaxies whose types were determined using the type values from the DR5 {\tt Photoz} table.}
\label{fig:s3ztype}
\end{figure}

The DR5 {\tt PhotoZ} table contains estimates of spectral type, given by a number between 0 and 1, where 0 is most red and 1 is most blue.  It has been shown (e.g., \citealt{Bud03}) that splitting the sample of galaxies into early- and late-type galaxies at a type value of 0.3 (hereafter denoted the {\it photo} sample) is roughly equivalent to splitting them using the \cite{Stra01} color criteria (hereafter denoted the {\it color} sample).  To test this, we split the entire sample into early- and late-type galaxies by photo and compare the resulting correlation functions to those measured for galaxies split by color.  The results for $s_3$ are displayed in Figure \ref{fig:s3ztype}.  While the two cuts produce similar results, there are some significant differences.  It appears that the characteristic rise at small scales for late-type galaxies is actually more prevalent in the galaxies split by the photo method.  This suggests that type confusion when splitting the galaxies by the color method (which would be expected to be greatest at higher redshift) may dampen the overall signal of late-type galaxies selected this way.  To minimize any type confusion, therefore, when we split our galaxy sample by redshift, their types will be determined via the photo method, as these types should be less sensitive to redshift than the color method (see \S \ref{sec:com} for further justification).     

\section{Higher-Order Correlations in Volume Limited Samples}
\subsection{ $z < 0.3 $}
The volume limited sample $z3$ contains roughly four million galaxies with $M_r < -19.5$.  We split this sample into fifteen different subsamples (though each is not mutually exclusive).  The sample is split by magnitude into two groups:  $-20.5 < M_r < -19.5$ and $-21.5 < M_r < -20.5$, and by redshift into two groups:  $0 < z < 0.2$ and $0.2 < z < 0.3$.  Each group itself contains three groups: early-type, late-type, and all-types, where the split is done by using the photo method.  

\begin{figure}[hbtp]
\plotone{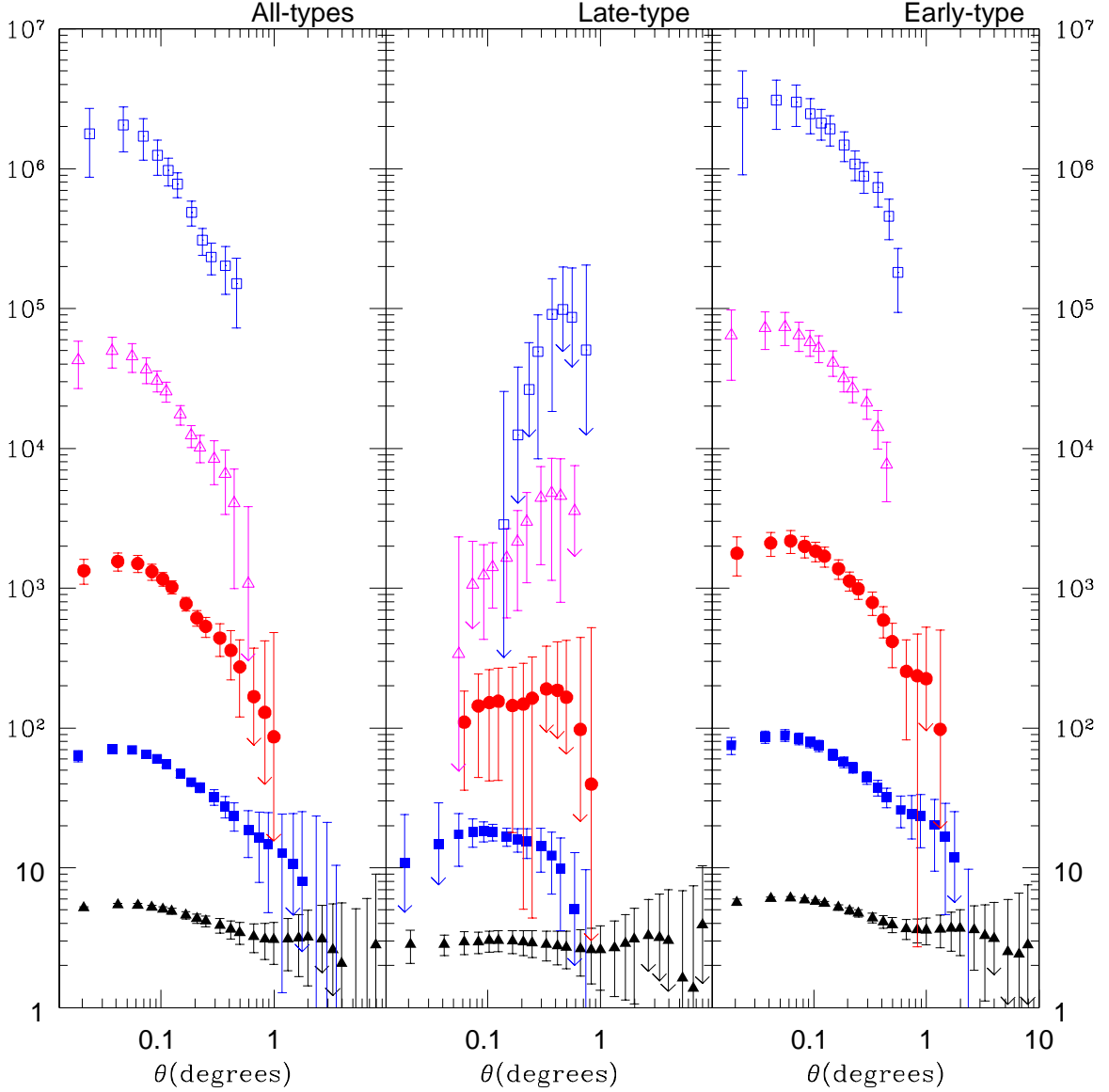}
\caption{The hierarchical amplitudes $s_{3}$ through $s_{7}$ for all-type (left), late-type (middle) and early-type (right) galaxies for the volume limited sample of galaxies $0 < z < 0.3$, $M_r < -19.5$. The late-type galaxies show significantly different behavior than the late-type measurement made using the full sample of galaxies.  The $\theta$ values for $s_{4}$ and $s_{7}$ have been shifted slightly and data with extremely large errors are omitted for clarity.  }
\label{fig:z.3all}
\end{figure}

Figure \ref{fig:z.3all} displays the $s_N$ measurements made using all-, late-, and early-type galaxies in the $z3$ volume limited sample.  Immediately, one notices that the late-type galaxies do not resemble the late-type galaxy measurements made on the full sample, especially at smaller scales.  The late-type galaxies also display larger error bars than the early-type galaxies, despite the fact that there are nearly the same number of late-type (1,984,021) galaxies and early-type galaxies (1,996,631) in this sample.  The early-type galaxies show significantly higher amplitudes than the late-type galaxies, and the difference is largest at $\sim$0.1$^0$, just as in the full sample of galaxies.

For each subset of galaxies (all, late, and early) the amplitudes for the full sample are slightly larger than those of $z3$.  This suggests that the bias is slightly larger in the $z3$ sample than in the full sample, as $s_N \propto b^{2-N}$.  This is not surprising, as the full sample contains less luminous galaxies than $z3$, and bias is known to increase with luminosity (e.g., \citealt{Ma03,Z05}).  The bias measurements for each sample are calculated in the following section.

\begin{figure}[hbtp]
\plotone{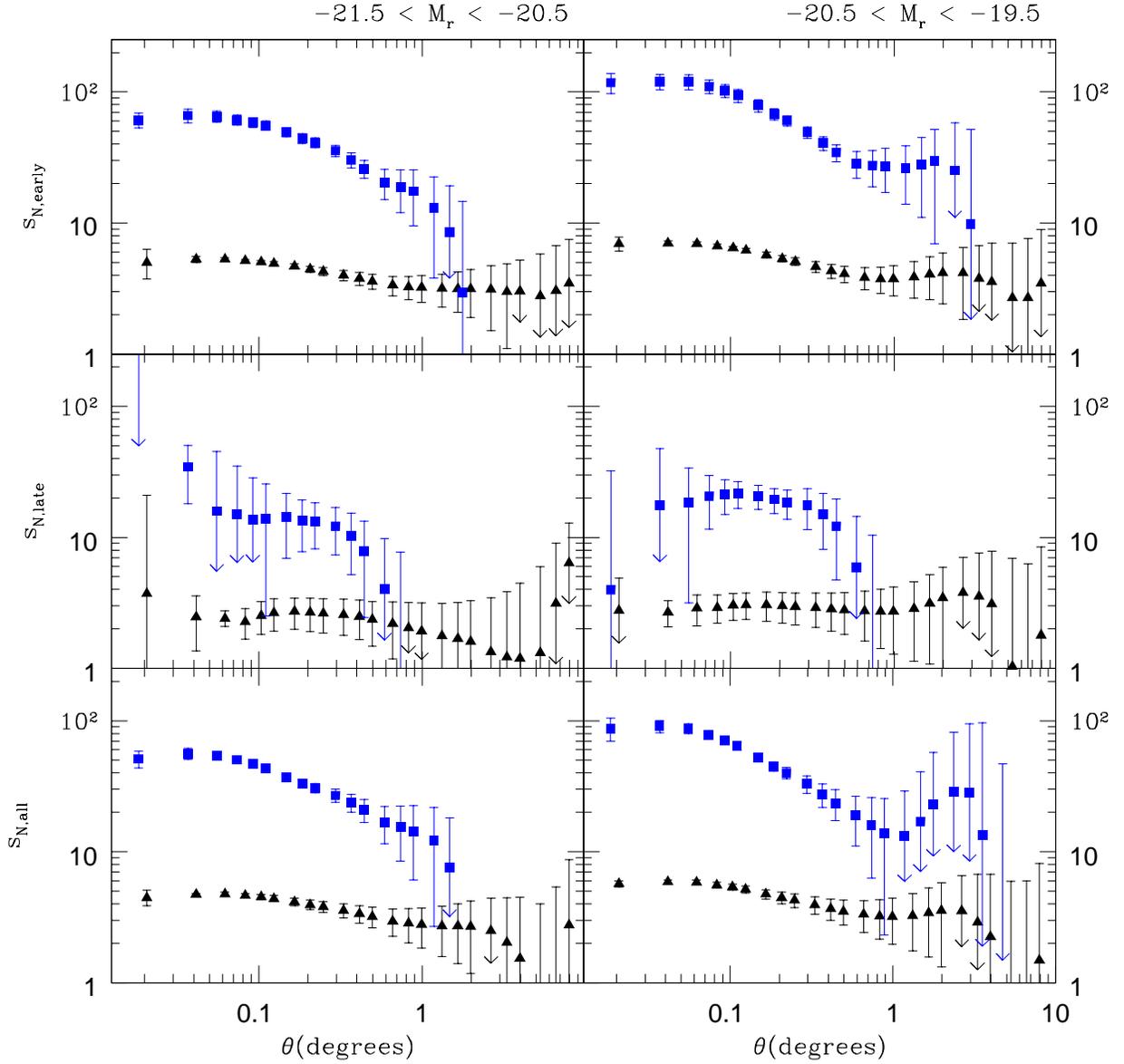}
\caption{The hierarchical amplitudes $s_{3}$ (black triangles) and $s_{4}$ (blue squares) for galaxies with $0 < z < 0.3$ split into $-21.5 < M_r < -20.5$ (left) and $-20.5 < M_r < -19.5$ (right) displayed for all- (bottom), late- (middle), and early-type (top) galaxies.  The $\theta$ values for $s_{4}$ have been shifted and data with extremely large errors are omitted for clarity (we also omit amplitudes $s_5$ through $s_7$ to be concise). The shapes are extremely similar, though the fainter galaxies display larger amplitudes, which is consistent with the more luminous galaxies having larger bias.}
\label{fig:z3Amag}
\end{figure}

Figure \ref{fig:z3Amag} displays the $s_3$ and $s_4$ measurements made by separating the $z3$ galaxies by luminosity into two groups, $-21.5 < M_r < -20.5$ and $-20.5 < M_r < -19.5$.  These measurements for all-, late-, and early-type galaxies are shown in the bottom, middle and top panels, respectively.  For each galaxy type, the shapes of $s_N$ for brighter galaxies are quite similar to the fainter galaxies.  The largest difference is that the amplitudes are higher for the fainter galaxies, which is consistent with the first-order bias increasing with luminosity.  Based solely on Figure \ref{fig:z3Amag}, it does not appear that non-linear clustering (i.e., $c_2$, $c_3$) is dependent on luminosity, as the differences in the amplitudes appear consistent with a linear bias model (see \S 6, Table \ref{T:bz3}). 

\begin{figure}[hbtp]
\plotone{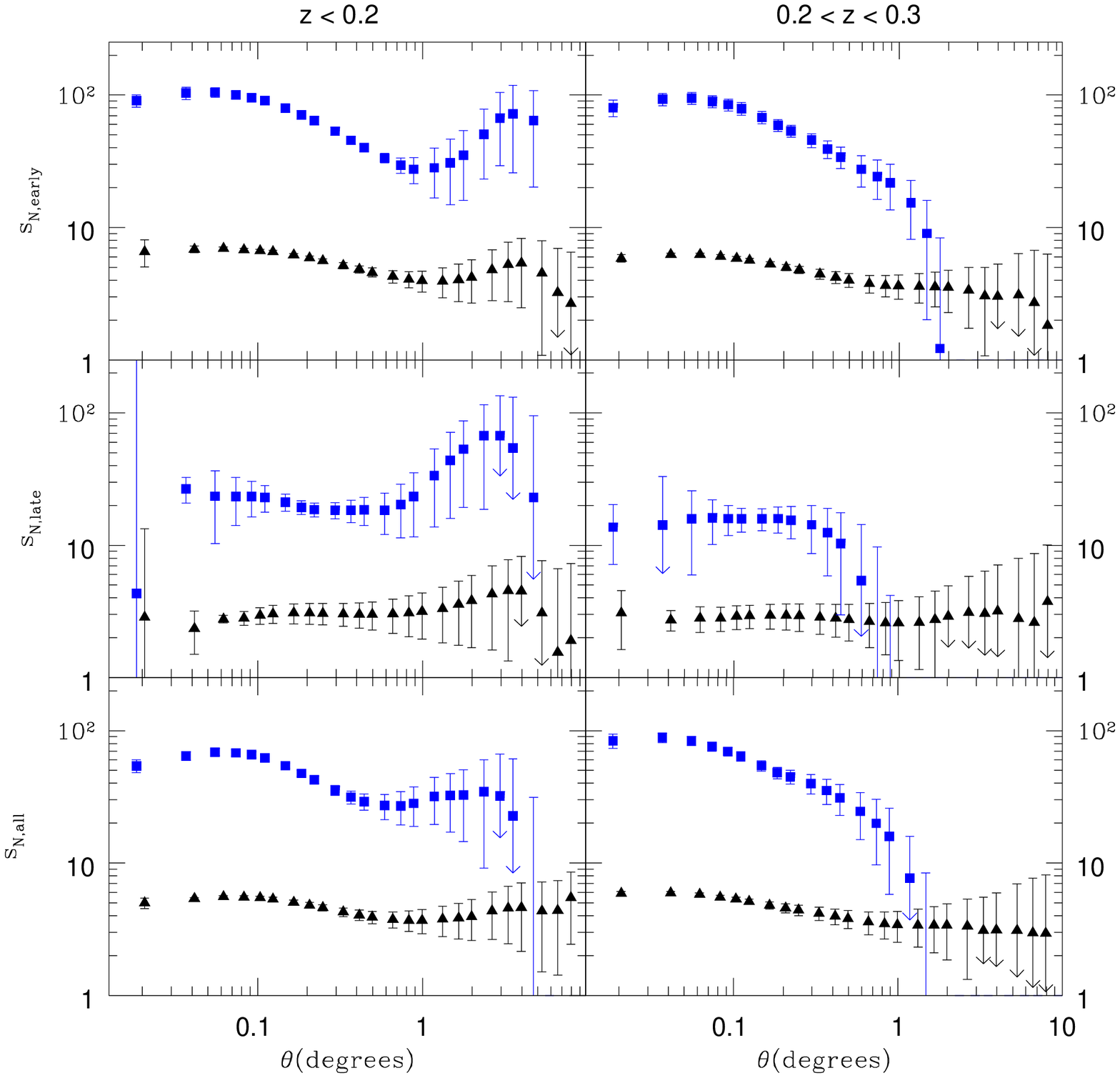}
\caption{The hierarchical amplitudes $s_{3}$ (black triangles) and $s_{4}$ (blue squares) for galaxies with $M_r < -19.5$ separated into two redshift bins, $0 < z < 0.2$ (left) and $0.2 < z < 0.3$ (right) displayed for all- (bottom), late- (middle), and early-type (top) galaxies.  The $\theta$ values for $s_{4}$ have been shifted and data with extremely large errors are omitted for clarity (we also omit amplitudes $s_5$ through $s_7$ to be concise).  }
\label{fig:z.2A}
\end{figure}

In order to test for evolution, $z3$ is split by redshift into a $0 < z < 0.2$ and $0.2 < z < 0.3$ sample.  Figure \ref{fig:z.2A} shows this split for all-, late-, and early-type galaxies in the bottom, middle and top panels, respectively.  After accounting for the fact that that the galaxies with $0.2 < z < 0.3$ probe scales that are about 1.6 times larger (based on median redshifts of 0.26 and 0.15) than the $0 < z < 0.2$ galaxies, the shapes of the $s_N$ are quite similar to a physical scale of about 5 $h^{-1}$ Mpc ($\sim$0.06$^0$ for $0 < z < 0.2$, $\sim$0.04 for $z < 0.3$).  Despite the visual differences, there does not appear to be significant evolution in the bias, as the differences in the measurements can be explained by the differences in the physical scales.

\begin{figure}[hbtp]
\plotone{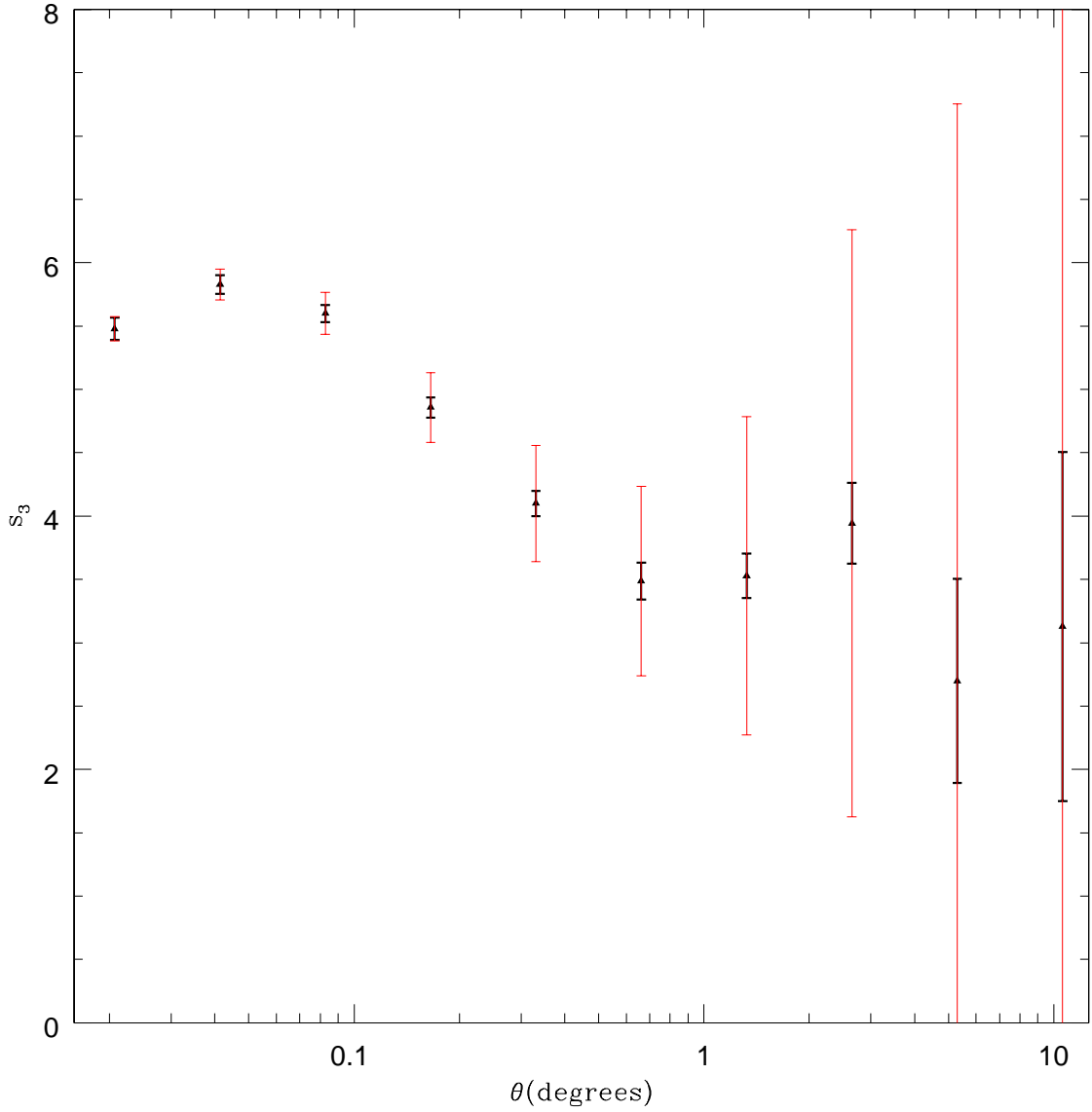}
\caption{The average $s_3$ measured from ten separate $z3$-like samples with redshifts determined by sampling the PDF of each galaxy's photometric redshift.  The bold black error bars represent the standard deviation of the ten measurements, while the light red error-bars represent the jackknife errors of the $s_3$ measurements using the true $z3$ sample.}
\label{fig:zerr}
\end{figure}

One concern with the volume limited samples we create is the possible effects of photometric redshift errors on the creation of our samples.  To quantify any potential bias, we created ten separate samples where instead of taking the stated photometric redshift of each galaxy, we sampled its probability-density-function (PDF) given by its one $\sigma$ error, thereby assigning a new redshift for each galaxy.  This allowed us to created ten separate samples with the same magnitude and redshift limits as $z3$.  Figure \ref{fig:zerr} displays the average measured $s_3$ of these samples, with errors (in bold black) calculated by finding the standard deviation of the ten measurements.  Underneath these points, the jackknife errors of the $z3$ sample are plotted in light red.  It is clear that the normal jackknife errors dominate the error budget, especially at larger scales where the errors are most important for fitting bias values.  Therefore, we conclude that photometric redshift errors should not make any significant difference in our results.

\subsection{ $z < 0.4 $}
\begin{figure}[hbtp]
\plotone{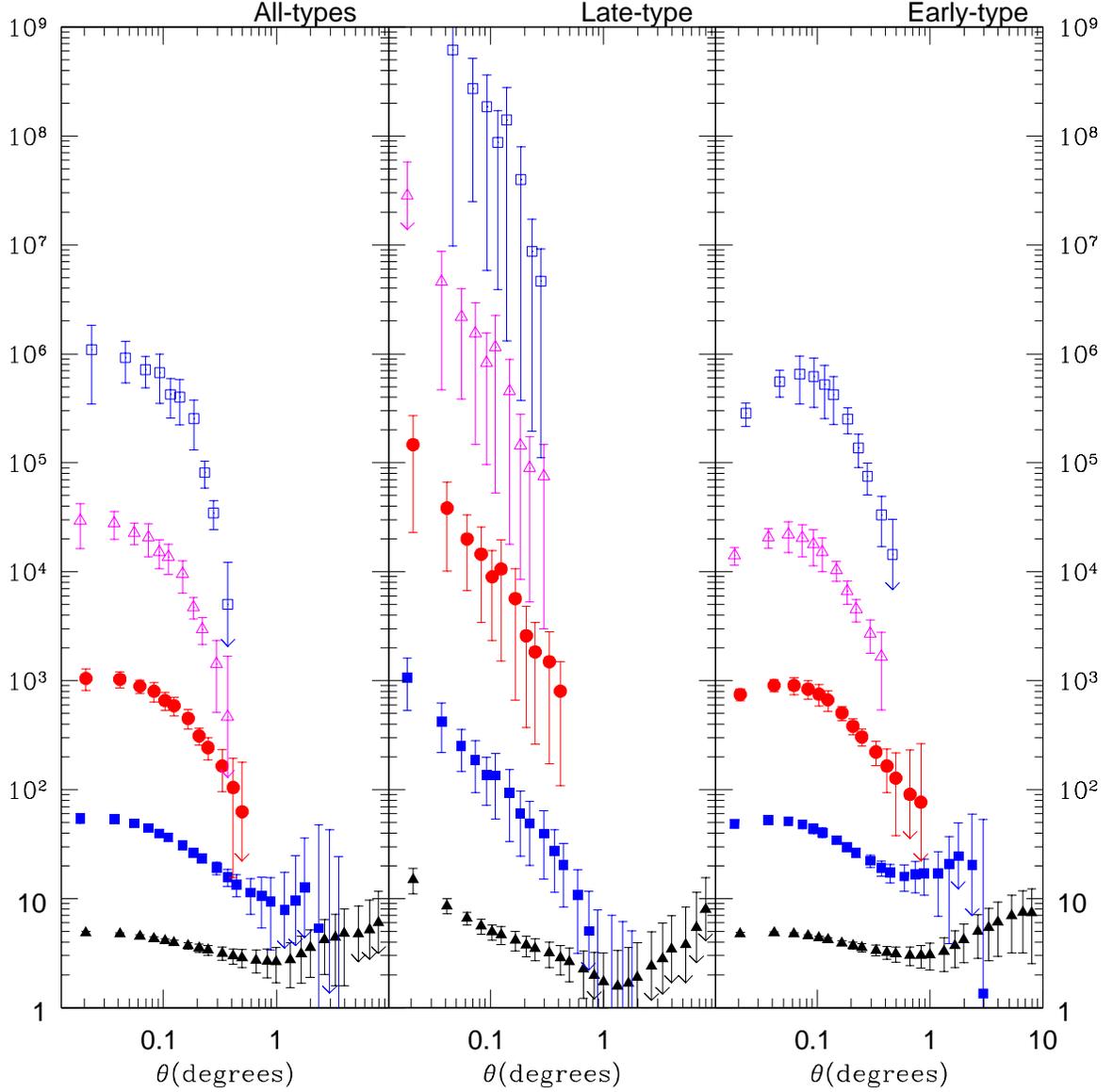}
\caption{The hierarchical amplitudes $s_{3}$ through $s_{7}$ for all-type (left), late-type (middle) and early-type (right) galaxies for the volume limited sample of galaxies $0 < z < 0.4$, and $M_r < -21.5$. These measurements look quite similar to those made on the full sample.  The $\theta$ values for $s_{4}$ and $s_{7}$ have been shifted slightly and data with extremely large errors are omitted for clarity.  }
\label{fig:z4all}
\end{figure}

The volume limited sample $z4$ contains just under three and a half million galaxies with $M_r < -20.5$ and $z < 0.4$.  Figure \ref{fig:z4all} shows the $s_N$ measurements for all-, late-, and early-type galaxies in $z4$, split using the photo method. Their shapes are quite similar to the measurements made on the full sample.  The $z4$ sample, however, shows a strong rise at small-scales in $s_N$ for late-types, unlike the $z3$ sample.  The error bars on the late-type galaxies are again larger that those of the early-type galaxies, but in this case it can be explained by the fact that there are significantly fewer late-type galaxies (1,325,488 late-types; 2,055,065 early-types).  

\begin{figure}[hbtp]
\plotone{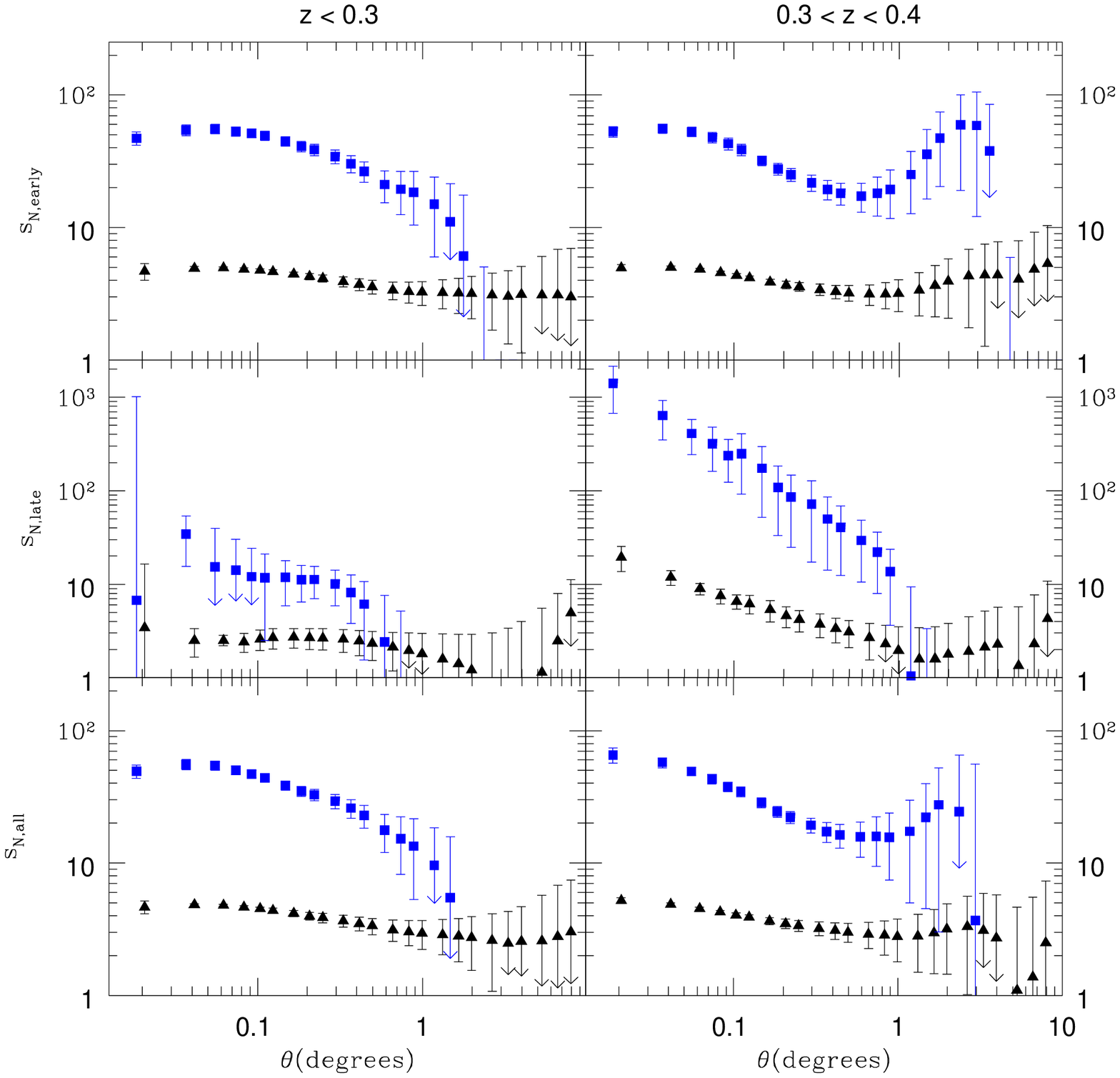}
\caption{The hierarchical amplitudes $s_{3}$ (black triangles) and $s_{4}$ (red squares) for galaxies with $M_r < -20.5$ separated into two redshift bins, $0 < z < 0.3$ (left) and $0.3 < z < 0.4$ (right) displayed for all- (bottom), late- (middle), and early-type (top) galaxies.  The $\theta$ values for $s_{4}$ have been shifted slightly and data with extremely large errors are omitted for clarity (we also omit amplitudes $s_5$ through $s_7$ to be concise).}
\label{fig:z.3.4A}
\end{figure}

Splitting $z4$ into two redshift bins, $z < 0.3$ (1,302,750 galaxies) and $0.3 < z < 0.4$ (2,077,803 galaxies) produces extremely interesting results.  Figure \ref{fig:z.3.4A} shows the $s_N$ measured in these redshift bins for all-, late-, and early-type galaxies in the bottom, middle, and top panels, respectively.  For each galaxy-type, the $s_N$ are much closer to constant at the lower redshift.  For the late-type galaxies, the errors are much larger at small scales for the low redshift bin.  This is due to the fact that there are nearly 75\% more high-redshift late-type (843,527) galaxies than low redshift (481,961).  Low redshift, late-type galaxies have a weak signal at small scales, while the high redshift galaxies have a relatively strong signal, meaning that, in the full sample, the high redshift signal dominates.  This explains the small-scale features seen in the full $z4$ sample of late-type galaxies.

\begin{figure}[hbtp]
\plotone{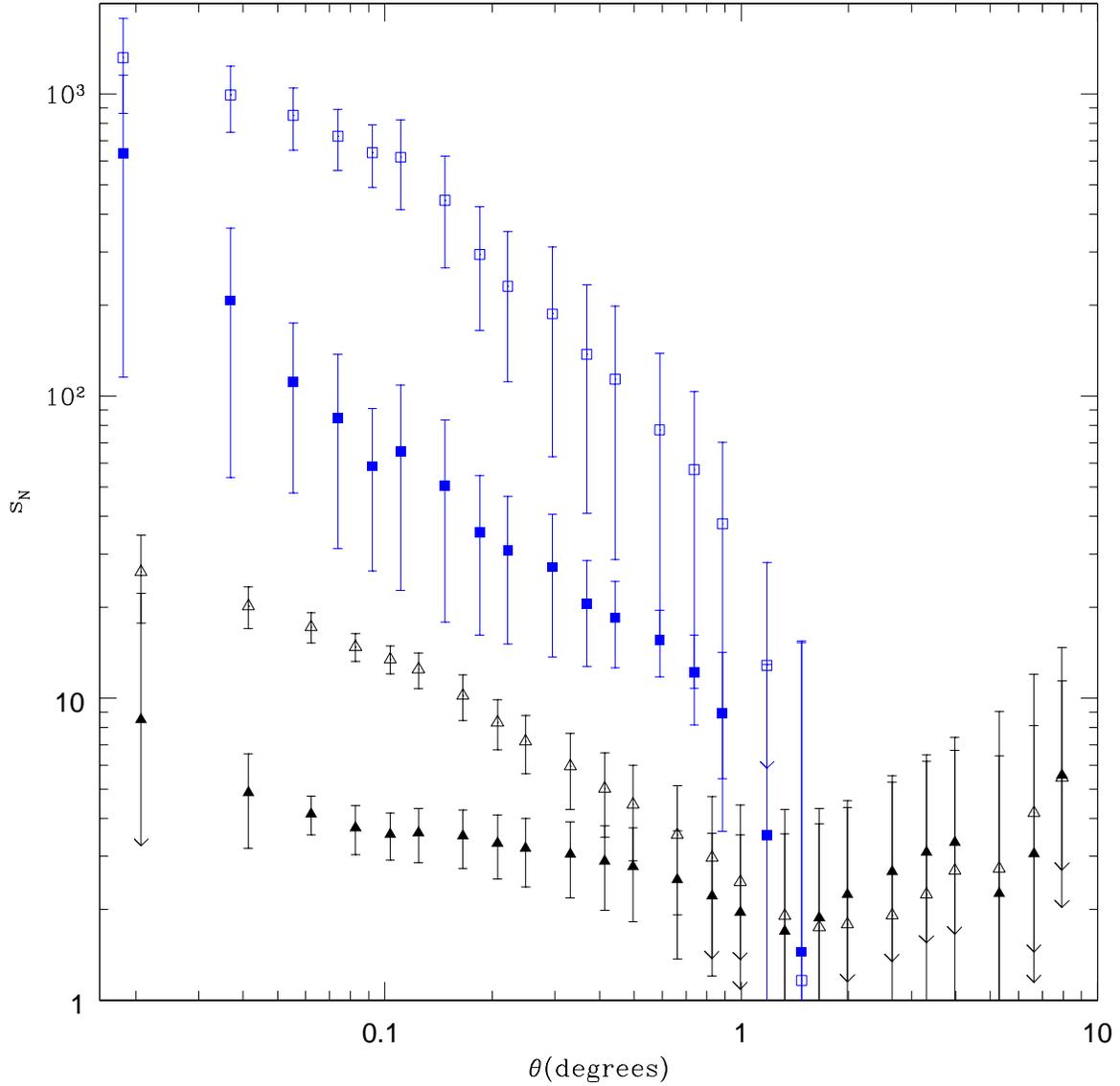}
\caption{The hierarchical amplitudes $s_{3}$ (black triangles) and $s_{4}$ (blue squares) for galaxies with $M_r < -20.5$ and $0.3 < z < 0.4$ displayed for $L1$ (closed symbols) and $L2$ (open symbols) galaxies.  The $\theta$ values for $s_{4}$ have been shifted slightly and data with extremely large errors are omitted for clarity.}
\label{fig:L2}
\end{figure}

The late-type galaxies can be further split by their type.  Separating the galaxies with type values greater than 0.3 at a type value of 0.65 produces a sample of galaxies that corresponds roughly to late-type spirals ($0.3 < t < 0.65$; denoted $L1$ hereafter) and one that corresponds roughly to irregular galaxies ($0.65 < t$; denoted $L2$ hereafter; \citealt{Bud03}).  Figure \ref{fig:L2} shows the measured $s_3$ and $s_4$ for $L1$ (closed symbols) and $L2$ (open symbols) for $0.3 < z < 0.4$.  The rise in the amplitudes at small scales is much stronger for the $L2$ galaxies,  which we interpret as evidence that the small scale rise correlates with star formation in dense environments, which we discuss further in \S \ref{sec:LT}.

\subsection{Comparison of Full Sample to Volume Limited}
\label{sec:com}
By comparing the $s_N$ measurements made on the full sample to those made on  the volume limited samples, features in the full sample of galaxies are isolated in the redshift/luminosity plane.  The most obvious conclusion is that the measurements of late-type galaxies are dominated by galaxies with $z > 0.3$, especially on angular scales smaller than half a degree, as the $s_N$ signal is much stronger at $z > 0.3$ for late-types.  For early-types, the shapes of $s_N$ measured in all subsamples are consistent both with each other and with the measurement of early-type galaxies drawn from the full sample.

\begin{figure}[hbtp]
\plotone{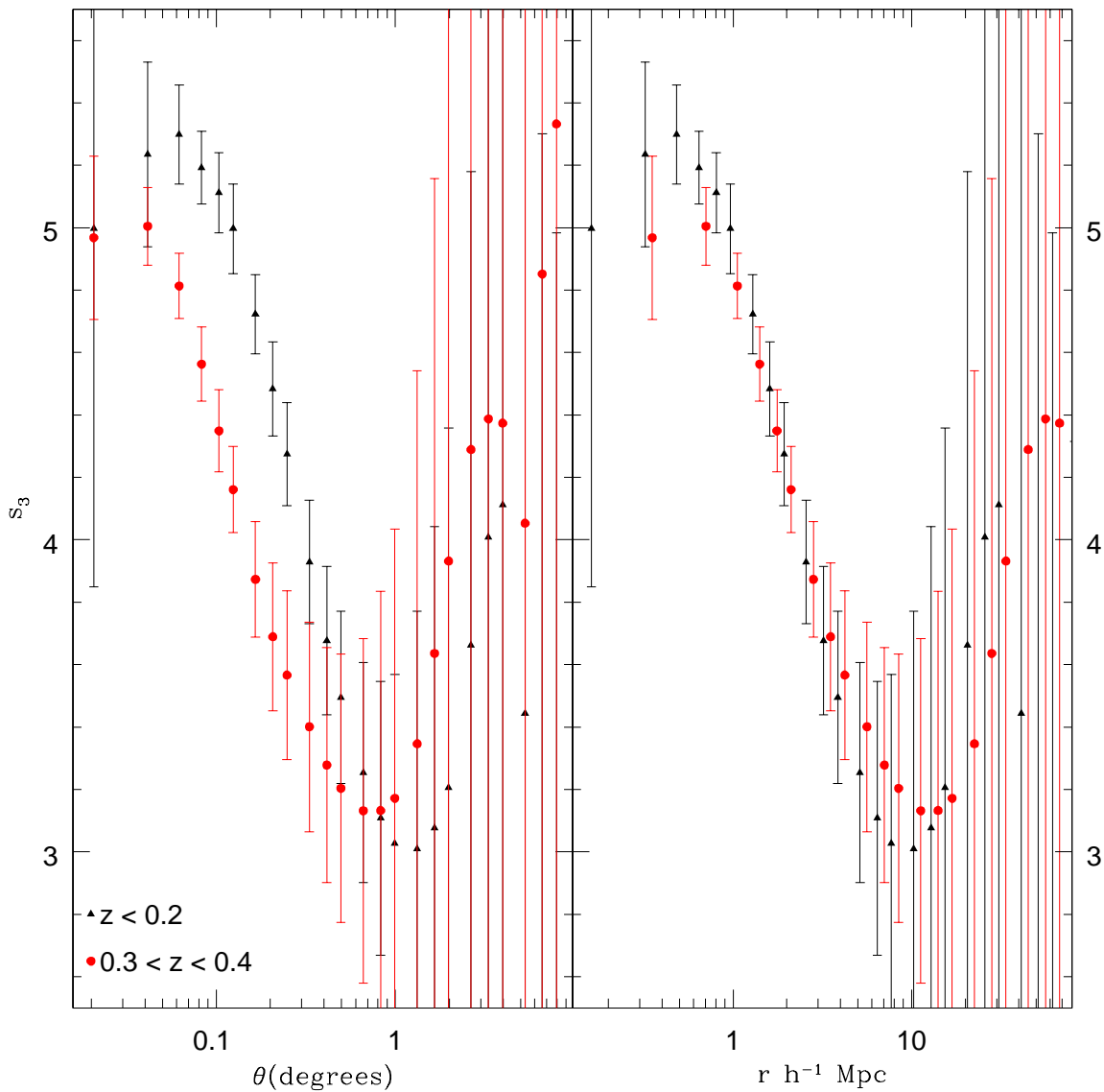}
\caption{The $s_3$ measurements for Early-type galaxies from the $z < 0.2, M_r < -19.5$ (black triangles) and the $0.3 < z < 0.4, M_r < -20.5$ (red circles) samples plotted against the angular scale (left panel) and the equivalent physical scale (right panel), with the lower redshift galaxies' $s_3$ divided by the relative bias of the two samples (1.31).  When plotted on the physical scale, the shape of the measurements agree, indicating that the features measured are physical in nature, and not a systematic effect. }
\label{fig:s3align}
\end{figure}

For many of the subsamples, there appears to be a feature at $\sim$2$^0$.  This is especially true for all galaxy types in the main sample of galaxies and for early-type galaxies with $0.3 < z < 0.4$ and $z < 0.2$.  It is quite prevalent for all galaxy types with $z < 0.4$ and is strangely not present for galaxies with $0.2 < z < 0.3$.  A close inspection of the feature reveals that it actually occurs at slightly different angular scales in the different redshift shells.  Figure \ref{fig:s3align} shows $s_3$ measured for early-type galaxies in the $z < 0.2$ redshift bin from $z3$ (red circles) and the $0.3 < z < 0.4$ redshift bin of $z4$ (black triangles).  These are the ideal samples to compare, as they represent the largest difference in physical scale.  The left panel shows the two measurements plotted on an angular scale, while the right panel plots $s_3$ against the equivalent physical scale (based on their median redshifts of 0.154 and 0.355).  When the two measurements are plotted against their equivalent scale, the match is much better.  This suggests that the feature is physical in nature and is characterized by a minima at 10 $h^{-1}$Mpc.  While this scale is interesting as it marks the transition from the non-linear to linear regime, the error bars preclude us from drawing any profound conclusions.

\begin{figure}[hbtp]
\plotone{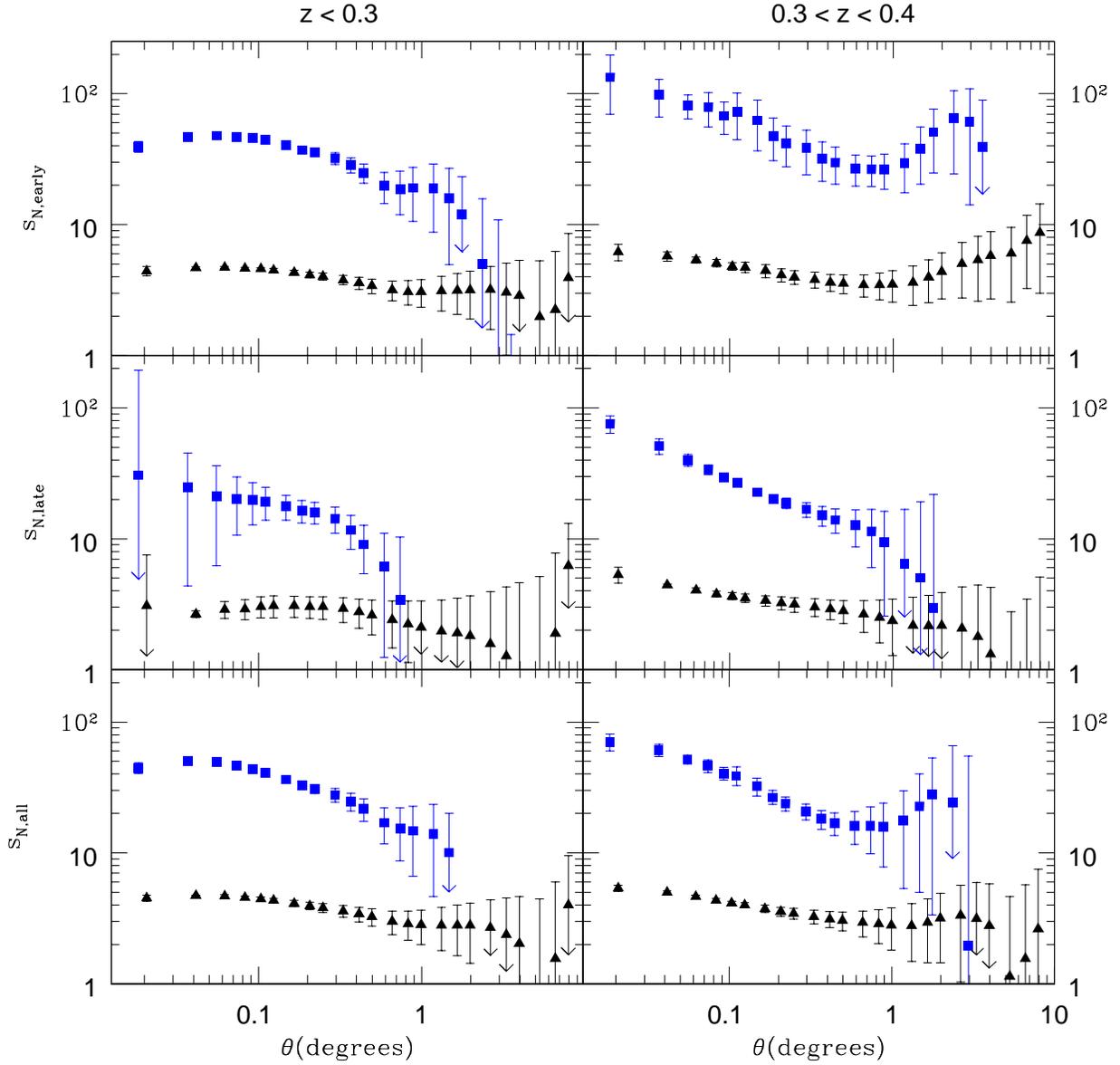}
\caption{Same as Figure \ref{fig:z.3.4A} only early- and late-type galaxies are split by the color method instead of the photo method.  This suggests that the color method does not adequately separate early- and late-type galaxies for $z > 0.3$.}
\label{fig:ELalt}
\end{figure}

Figure \ref{fig:s3ztype} suggested that separating early- and late-type galaxies via the color method may break down at higher redshifts.  Figure \ref{fig:ELalt} displays the same information as Figure \ref{fig:z.3.4A}, with galaxy-type determined via the color method instead of the photo method.  The measurements for $z < 0.3$ are nearly identical to those measured splitting by the photo method, but for $0.3 < z < 0.4$ the results are quite different, for both the early- and late-type galaxies.  This suggests that for $z > 0.3$, the color method cannot adequately distinguish between early- and late-type galaxies, and justifies our preference for separating galaxy type based on the photo method.

\section{Bias Measurements}
There is no guarantee that galaxies cluster with the same amplitude as the dark matter they trace.  Indeed, it has often been shown that galaxies display different clustering amplitudes when separated by luminosity or type (e.g., \citealt{Fry93,W98,N02,Ma03,Z05,Ross,Cr06}).  The simplest model for this difference is the linear bias model.  In this model, the overdensity of galaxies is a linear function of the overdensity of dark matter:
\begin{equation}
\delta_g = b\delta_{DM}
\end{equation}
where $g$ denotes galaxy, $DM$ denotes dark matter, and $b$ is the linear bias factor.  In this approximation, one finds the simple relationship that 
\begin{equation}
\bar{\omega}_{2,g} = b^2\bar{\omega}_{2,DM}.  
\label{eq:linb}
\end{equation}
It has been shown, however (e.g., \citealt{Fry93,Gaz05,Ross,Cr06}), that linear bias may not be a good approximation.  One can represent the relationship between overdensities more generally, such that the measured overdensity is some function of the dark matter overdensity.  As such, the relationship can be expanded into a Taylor series \citep{Fry93}:
\begin{equation}
\delta_{g} = \sum_{N=0}^{\infty}{\frac{b_{N}}{N!}\delta^{N}_{DM}}
\label{eq:bias}
\end{equation} 
where $b_N$ is the $N$th order bias term.  Thus it follows that if non-linear bias is important, a measurement of the linear bias will increase as the overdensities increase (i.e., as the scale gets smaller) and the non-linear terms (i.e., any $b_N$ for $N > 1$) grow in importance.  When $\delta \ll 1$,$~b_1$ can be determined via Equation \ref{eq:linb}, as higher order terms will be negligible.  

\subsection{First-Order Bias Measurements\label{sec:B1}}
\begin{figure}[hbtp]
\plotone{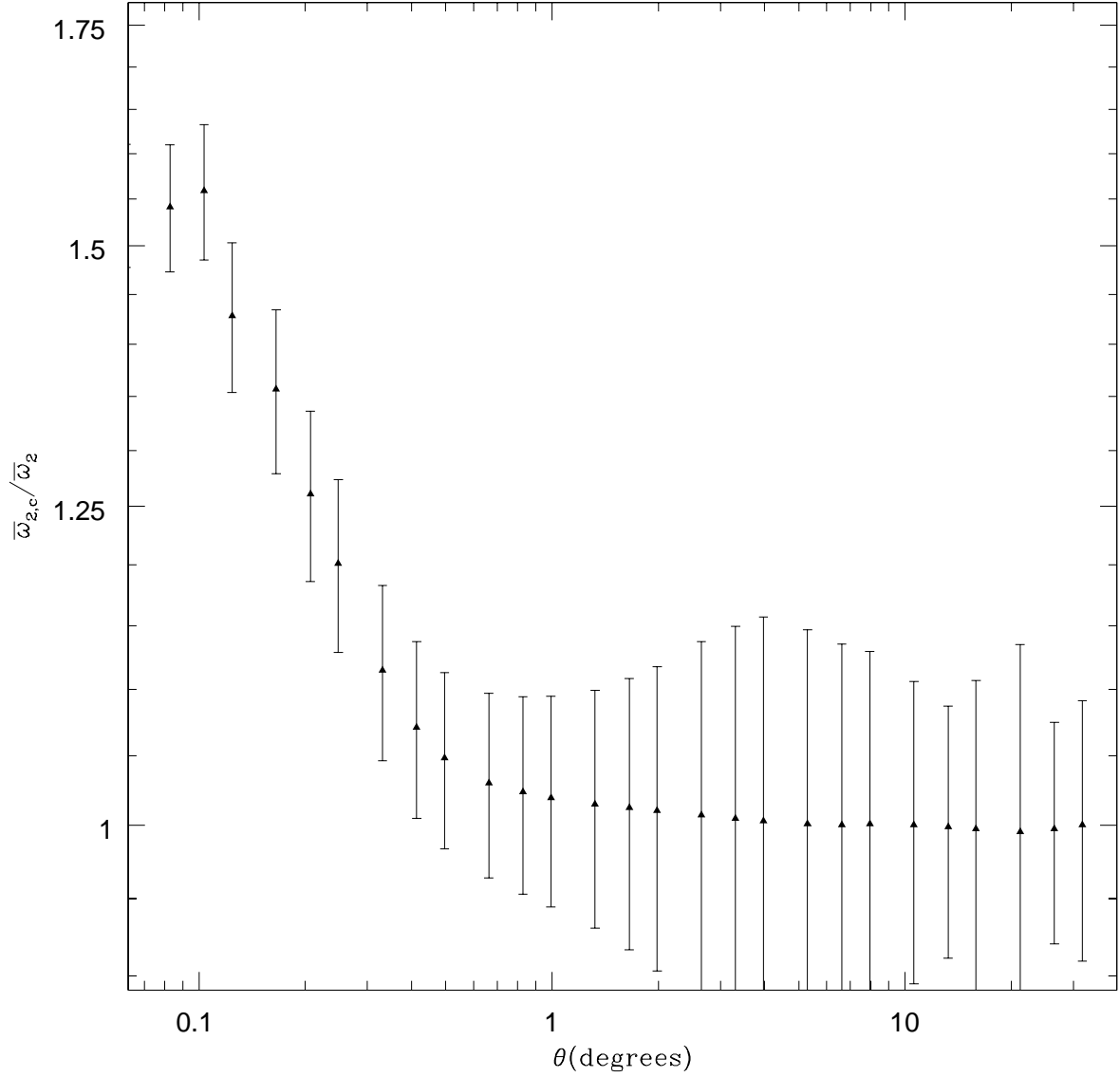}
\caption{The ratio of $\bar{\omega}_{2,c}$ to $\bar{\omega}_2$, where $\bar{\omega}_{2,c}$ calculated by correcting over-densities for a second-order bias term ($b_2$) equal to -0.3 and $\bar{\omega}_2$ is calculated in the standard way.  Both are calculated using the $z3$ sample of galaxies with $-20.5 < M_r < -19.5$.  The ratio begins to grow significantly greater than 1 for $\theta < 0.66^{0}$, corresponding to a physical scale of approximately 8$h^{-1}$Mpc.  This corresponds with the the lower bound of the weakly non-linear regime.  All bias values are thus calculated for $r > 8h^{-1}$Mpc.}
\label{fig:w2bi}
\end{figure}

In order to find $b_1$, one must first determine the scales at which higher-order terms make a negligible contribution to $\bar{\omega}_2$.  This can be done by calculating $\bar{\omega}_{2}$, incorporating a second-order bias term, and comparing it to the $\bar{\omega}_{2}$ calculated normally.  To second order, the dark matter over-density can be expressed (via trivial manipulation of Equation \ref{eq:bias}) as
\begin{equation}
\delta_{DM} = b^{-1}_2\left(-b_1 \pm \sqrt{b^2_1 + 2b_2\delta_g} \right)  
\label{eq:dcorr}
\end{equation}
Thus, when calculating $\bar{\omega}_2$, if Equation \ref{eq:dcorr} is used to correct the measured overdensities in each cell, a measurement is returned that assumes a first and second order bias.  If the first-order bias is set to 1 and a reasonable second-order bias is applied, the resulting measurement will begin to deviate significantly from the standard measurement when the second order effects become important.  Based on previous measurements \citep{Gaz05,Ross}, we select -0.3 as a reasonable value to use.  Figure \ref{fig:w2bi} shows the ratio of $\bar{\omega}_{2}$ calculated with this value to $\bar{\omega}_{2}$ calculated in the standard way, using the sample of galaxies from $z3$ with $-20.5 < M_r < -19.5$.  The ratio grows significantly greater than 1 for $\theta < 0.66^{0}$, corresponding to a physical scale of about 8$h^{-1}$Mpc.  This is right where one would expect non-linear effects to become important, as it marks the generally accepted transition to the weakly linear regime.  As a result, we calculate the first order bias for scales where the corresponding physical scale ($\bar{r}$) is greater than 8$h^{-1}$Mpc.     

Calculating the bias requires knowing $\bar{\omega}_{2,DM}$.  Of course this is not directly measurable, so we must resort to using a theoretical model.  \cite{Smith} have derived fitting formulae based on $N$-body simulations, which produce matter power-spectra when given specific input cosmological parameters.  By using a modified version of Limber's equation \citep{P80}, one can use the appropriate redshift distribution to invert the $P(k)$ to obtain $\bar{\omega}_2(\theta)$:
\begin{equation}
\bar{\omega}_2(\theta) = \pi\int \left(\frac{dn}{dz}\right)^2\left(\frac{dZ}{d\chi}\right)F(\chi)dz\int P(k)W^2_{2D}(D\theta k)dk
\end{equation}
\noindent where $W_{2D} = 2\frac{J_{1}(x)}{x}$ is the top-hat two-dimensional window function, $D$ is the survey depth (determined by the median redshift), $P(k)$ is the matter power spectra, $k$ is the spectral index, and $J_1$ is the first-order Bessel function of the first kind.  In a flat universe, $F(\chi) = 1$, and $dz/d\chi = H(z)/c = H_0\sqrt{\Omega_m(1+z)^3 + \Omega_\Lambda/c}$, simplifying Equation 10 to:
\begin{equation}
\bar{\omega}_2(\theta) = \frac{H_0\pi}{c}\int (\frac{dn}{dz})^2\sqrt{\Omega_m(1+z)^3+\Omega_\Lambda}dz\int P(k)W^2_{2D}(D\theta k)dk
\end{equation}
This integral was numerically determined with our assumed cosmology and fixing the value of $\sigma_8$ for the matter $P(k)$ to 0.8.  Following this procedure, model $\bar{\omega}_2$ were produced for each subsample studied, using the appropriate redshift distribution.  Finally, the first order bias was calculated by using the covariance matrix at scales $\bar{r} > 8h^{-1}$Mpc.  We note that this approach is much more accurate than the methods employed in Ross06, which inverted the measurements to real-space, required $\bar{\omega}_{2}$ be a power-law, calculated only the relative bias, and were insensitive to any changes in the bias as a function of scale.

Tables \ref{T:bz3} and \ref{T:bz4} display our calculated $b_1$ values for all of the measurements made using the volume-limited samples $z3$ and $z4$.  The most important information can be summarized as follows:  (1) for $z3$, there is no significant evolution in $b_1$ as a function of redshift for any galaxy type, but there is significant evolution seen in $z4$ and it is most dramatic for the late-type galaxies;  (2) $b_1$ grows larger with luminosity, independent of galaxy type; and (3) $b_1$ is consistently larger for early-type galaxies, though the ratio of $b_{1,early}$ to $b_{1,late}$ is larger for $z < 0.3$.  The determined luminosity and type dependences of $b_1$ are consistent with the general results of previous findings (\citealt{Fry93,W98,N02,Ma03,Z05,Ross,Cr06}).  All of the determined values are dependent on the true value of $\sigma_8$.  Since all of the models use $\sigma_8 = 0.8$, the true values of the $b_1$ are $0.8 b_{1,m}/\sigma_8$, where $b_{1,m}$ are the measured values reported in Tables \ref{T:bz3} and \ref{T:bz4}.

Most of the fits to $b_1$ using the $z3$ sample minimize $\chi^2$ such that it is approximately one or smaller per degree of freedom (DOF), meaning most fits favor a scale invariant $b_1$ over the fitted range.  The notable exception is the measurement for late-type galaxies with $0.2 < z < 0.3$, with a $\chi^2$ per DOF value of nearly two.  The fits to $b_1$ for $z4$ are in general worse than those of $z3$.  All but two have fits with $\chi^2$ per DOF greater than one.  It is thus unlikely that $b_1$ is scale invariant at higher redshifts, a fact which must be taken into account when calculating higher-order bias terms.  

\subsection{Second-Order Bias Measurements\label{sec:HOB}}
In order to measure the second order bias, we use the equation \citep{Fry93}:
\begin{equation}
S_{3,T}=b_{1,T}^{-1}(S_{3,DM} + 3c_{2,T})
\end{equation}
where $c_2 = \frac{b_2}{b_1}$.  This equation was derived for real-space hierarchical amplitudes, but given a theoretical $s_{3,DM}$, $c_2$ can be determined in the same way from $s_{3,g}$.  Using PT, an expression valid in the weakly non-linear regime can be derived for $\bar{\omega}_{3,DM}$ \citep{Ber95}:
\begin{equation}
\begin{array}{l}
\bar{\omega}_{3,DM} = 6\left(\frac{H_0\pi}{c}\right)^2\int\left(\frac{dn}{dz}\right) ^3\left[\Omega_m(1+z)^3+\Omega_\Lambda\right]dz~\times\cr \\  \left\{\frac{6}{7}\left(\int kP(k)\left[W^2_{2D}(D\theta k]\right)^2dk\right) + \nonumber\int kP(k)\left(W^2_{2D}(D\theta k)\right)dk\int k^2D\theta P(k)W_{2D}(D\theta k)W^{'}_{2D}(D\theta k)dk\right\} 
\end{array}
\end{equation}
From Equation 3, it is clear that $s_{3,DM}$ can be obtained by dividing this $\bar{\omega}_{3,DM}$ by $\bar{\omega}^2_{2,DM}$.  Equation 13 uses PT, which means that a linear power spectrum must be used when calculating $s_{3,DM}$, and it is only valid at scales with $r \gtrsim 8~ h^{-1}$Mpc.  Using the model $s_{3,DM}$ and $b_1$ calculated at each scale (allowing for any scale dependence in $b_1$), one can calculate $c_2$ at each scale and thus construct its covariance matrix and determine the $\chi^2$ best-fit average values (see \S \ref{sec:err}) of $c_2$ for each subsample.  Once again, we note that this method is superior to the one employed by Ross06, as it has been shown (e.g. \citealt{Gaz98,Ber02}) that this method is a better match to simulations than assuming a hierarchy and inverting $s_N$ to real-space.

The values of $c_2$ for the $z3$ and $z4$ samples are presented in the right hand columns of Tables \ref{T:bz3} and \ref{T:bz4}.  The important aspects are as follows: (1) as first seen in Ross06, the $c_2$ values for late-type galaxies are significantly lower than those of early-type galaxies; (2) early-type galaxies have a value of $c_2$ that is independent of both luminosity and redshift; and (3) $c_{2,late}$ varies slightly with luminosity and evolves significantly between redshifts of 0.3 and 0.4, which we explore in \S \ref{sec:LT}.  The $\chi^2$ per DOF for all but three measurements are less than 1.24. The three exceptions are for measurements drawn from samples of early-type galaxies, and the poor fits are largely due to the strength in the feature at 10 $h^{-1}$Mpc.  To the limit on our data, one would expect most measurements can be fit by a constant $c_2$ , as the errors on the $s_3$ measurements over the fit ranges are typically large.

\section{Discussion}

\subsection{$c_2$ vs. $\sigma_8$}
\begin{figure}[hbtp]
\plotone{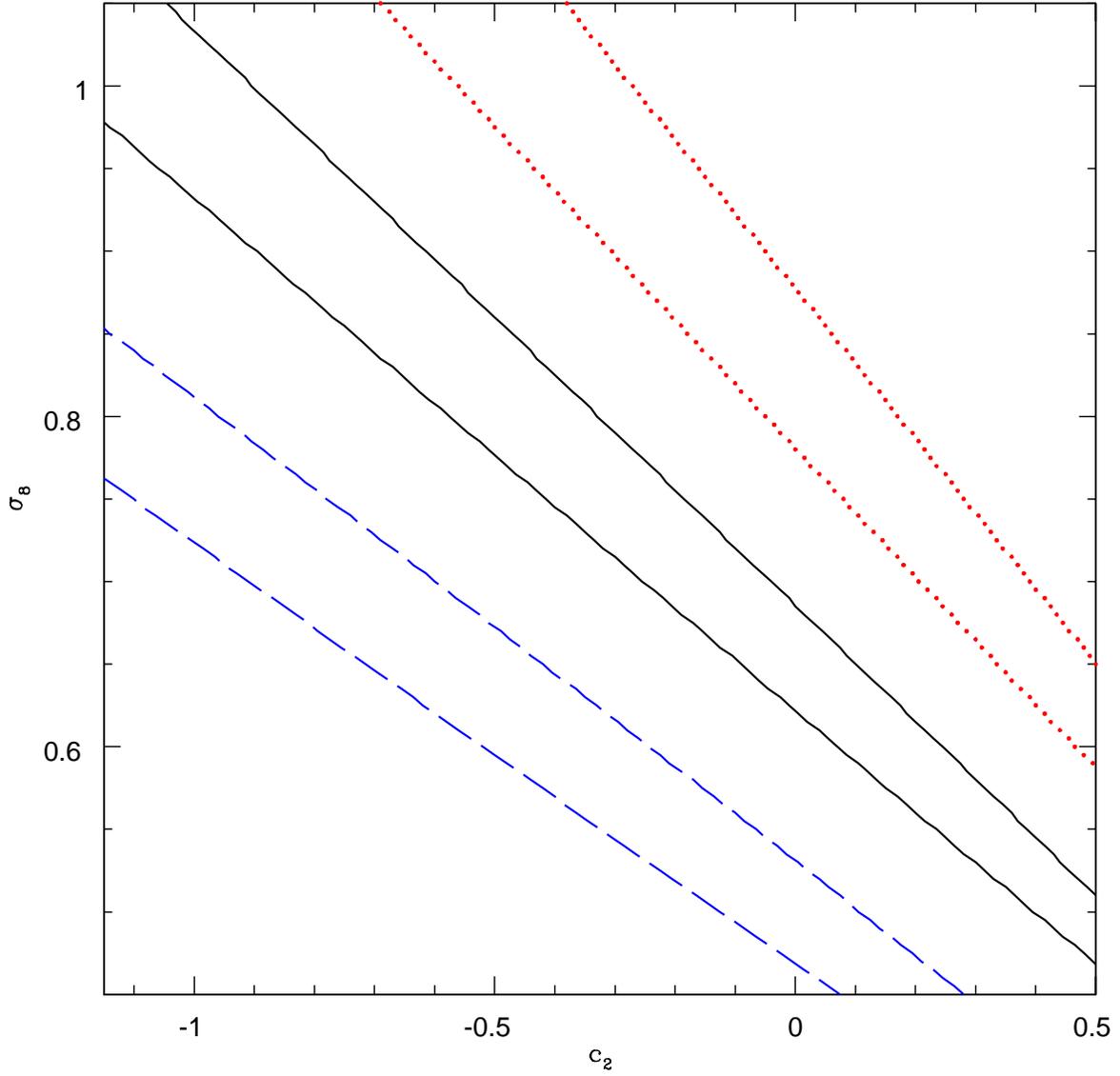}
\caption{The 1 $\sigma$ allowed regions of parameter space for $\sigma_8$ and $c_2$ for all- (black, solid), early- (red, dashed), and late-type (blue, dotted) galaxies.  The allowed second-order bias varies greatly with $\sigma_8$, but the differences in $c_2$ as a function of galaxy type are robust.}
\label{fig:sig8}
\end{figure}

The relationship between $c_2$ and $\sigma_8$ is more complicated than for the first-order bias.  Given that our measured $c_{2,m}$ are calculated using $\sigma_8 =  0.8$ for the matter $P(k)$, it can be shown that
\begin{equation}
c_2 = \frac{\sigma_8}{0.8}c_{2,m} + \frac{s_3}{3}\left(1-\frac{\sigma_8}{0.8}\right)
\end{equation}
where $c_2$ is the true $c_2$.  Since $s_3$ is not constant, the difference between $c_2$ and $c_{2,m}$ is not constant either.  Thus, it is possible to perform a two parameter fit on the values of $c_2$ and $\sigma_8$, thereby placing loose constraints on $\sigma_8$.  Figure \ref{fig:sig8} shows the one $\sigma$ allowed regions of parameter space for $c_2$ and $\sigma_8$ for all- (black), early- (red), and late-type (blue) galaxies in the $z3$ volume limited sample.  Clearly, there is a large spread in the allowed values of $c_2$ even if one constrains $0.7 < \sigma_8 < 0.9$ (changing $\sigma_8$ from 0.7 to 0.9 decreases $c_2$ by about 0.6 for all galaxies type).  It is clear that a significant difference in the $c_2$ of early- and late-type galaxies is robust against changes in $\sigma_8$, as increasing $\sigma_8$ has only a slight effect, (the difference is 1.0 for $\sigma_8 = 0.7$ and it is 1.1 for $\sigma_8 = 0.9$).  Thus, irrespective of $\sigma_8$, there is clear evidence for non-linear bias differences between early- and late-type galaxies.

Focusing on the $z3$ sample of all galaxy types, if we set $c_2 = 0$ we constrain the value of $\sigma_8$ to be in the range 0.64+0.04/-0.03. This is inconsistent at the one $\sigma$ level with the WMAP3 best-fit parameters ($\sigma_8 =0.74^{+0.05}_{-0.06}$) but consistent with the one $\sigma$ lower bound on WMAP3 as constrained by inflationary models (e.g, $\sigma_8=0.700^{+0.063}_{-0.065}$).  On the other hand, it is highly at odds with the typical effect of combining WMAP3 and large-scale structure constraints, which tend to higher values of $\sigma_8$.  We would contend that these higher values of $\sigma_8$ may be inconsistent with linear theory, as our data suggest $c_2$, for all galaxies, deviates by more than two $\sigma$ from zero for $\sigma_8 > 0.71$.  If we instead set $c_2 = -0.36$, as found by \cite{Gaz05}, $\sigma_8 = 0.77\pm0.04$ --- nearly identical to the WMAP3-SDSS combined measurement.  Thus, considering that it is both unlikely that $c_2$ is greatly negative and that it is unlikely that $\sigma_8$ is less than 0.68, we contend that the best interpretation of our results is that $c_2$ is at least slightly negative and that $\sigma_8 < 0.8$.

With the methods employed in this paper, it is not possible to break the degeneracy between the bias parameters and $\sigma_8$.  It may be possible, however, using measurements similar to those presented herein.  When we tested the effects of $b_2$ on the calculation of $\bar{\omega}_2$, we created a way in which one could compare a $\bar{\omega}_2$ measurement `corrected' for $b_2$ to a model $\bar{\omega}_2$ and thus test whether or not a particular value of $b_2$ was a good fit.  In principle, one could calculate $\bar{\omega}_2$, varying the input value of $b_2$, to find a $\chi^2$ best-fit value of $b_2$.  This best fit value of $b_2$ would be independent of $\sigma_8$, as $b_2$ affects the shape of $\bar{\omega}_2$, and $\sigma_8$ affects only the normalization.  Such an effort would require significant computational resources, and would certainly be prohibitive on the twenty-four separate subsamples we studied.  Endeavoring to such a task using a sample that probes large volumes with minimal shot noise and precise photometric redshift estimations, such as the luminous red galaxies (LRGs), is more feasible (Ross et. al. 2007, in preparation).

\subsection{$c_2$ vs. $b_1$}
Recent studies have found a relationship between $c_2$ and $b_1$.  \cite{Gaz05} suggested the relationship $c_2 = b_1 - 1.2$.  \cite{Nish06} measured results consistent with this relationship, and they showed that $c_2$'s dependence on $b_1$ is physically motivated.  The major difference between these studies and ours is that we use galaxies over a much larger redshift range.  Thus, we must determine if the theoretical relationship between $c_2$ and $b_1$ is robust against changes in redshift.

Following the methods of \cite{Nish06}, it is possible to calculate the first- and second-order bias parameters using a simple HOD model where the mean number of galaxies in a halo of mass $M$ is given by
\begin{equation}
\left<N(M)\right> = \left\{
 \begin{array}{ll}
 1 + \left(\frac{M}{M_1}\right)^{\alpha} & M > M_{min} \\
 0 & M \leq M_{min}
 \end{array}
 \right.
 \end{equation}
where $\alpha$, $M_1$, and $M_{min}$ are free parameters and in general are fit such that this HOD model reproduces the measured clustering.  Given this HOD model, the bias parameter $b_N$ is given by
\begin{equation}
b_N = \frac{\int^{M_{max}}_{M_{min}}dMn_{halo}(M,z)\left<N(M)\right>B_N(M,z)}{\int^{M_{max}}_{M_{min}}dMn_{halo}(M,z)\left<N(M)\right>}
\end{equation}
where $n_{halo}$ is the mass function of halos with mass $m$ at redshift $z$ determined via an ellipsoidal collapse model (e.g., \citealt{Sheth01}) and $B_n(m,z)$ is the $n$th-order bias coefficient of halos.  We calculate both factors following the methods described in detail in \cite{Nish06}, but with $\sigma_8 = 0.8$ so that we can compare with the calculated bias with our bias measurements (made assuming $\sigma_8 = 0.8$).  We adopt the fit parameters of \cite{Z05} determined for $M_r < -19.5$ (log$_{10}M_{min}$ = 11.76, log$_{10}M_{1} = 13.15$, $\alpha =1.13$) and $M_r < -20.5$ (log$_{10}M_{min}$ = 12.30, log$_{10}M_{1} = 13.67$, $\alpha =1.21$) for our HOD models which we use to calculate $b_1$ and $c_2$ at the median redshifts of the different subsamples.

\begin{figure}[hbtp]
\plotone{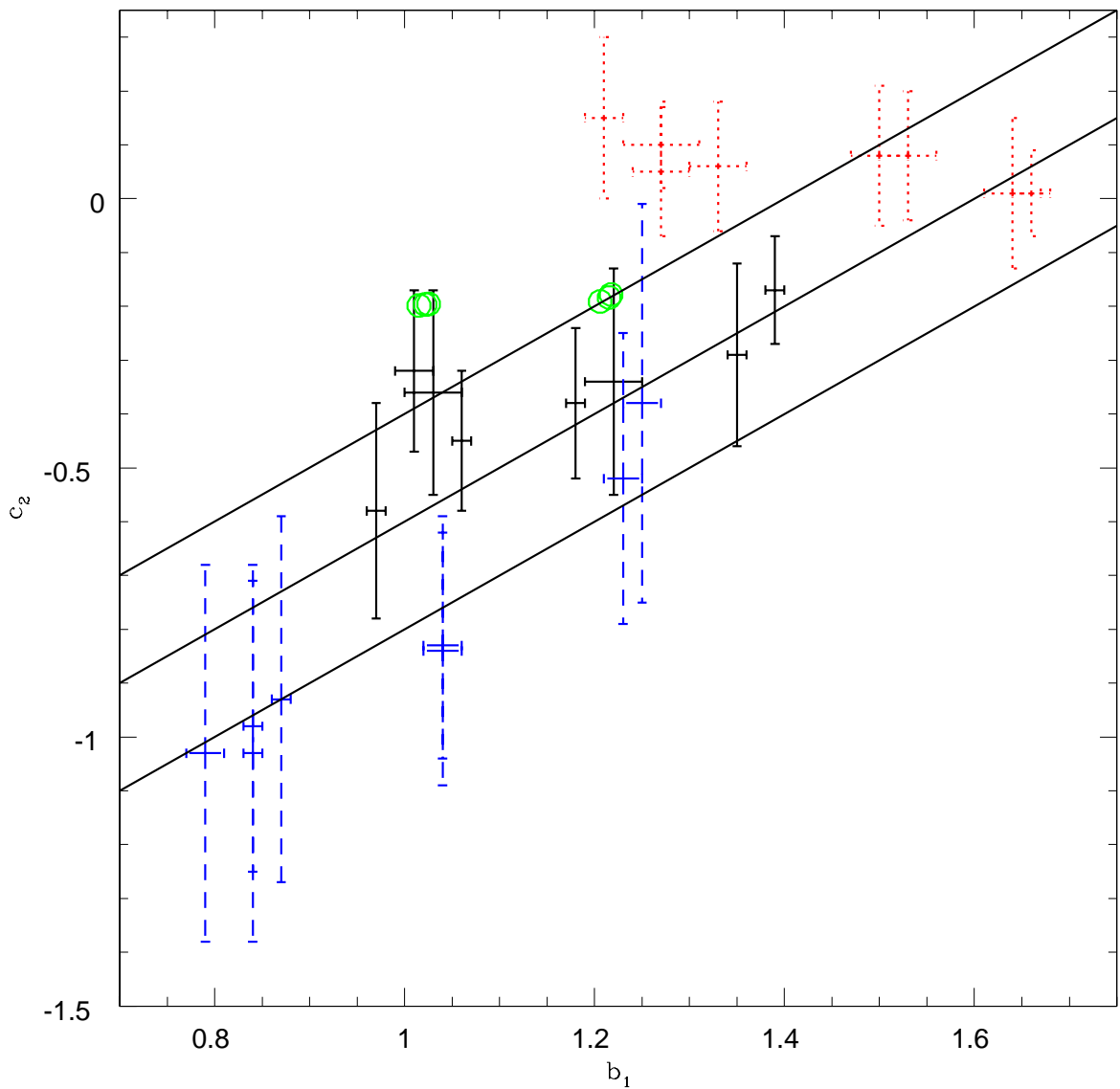}
\caption{The measured $c_2$ for each subsample plotted against the measured $b_1$.  Each point is marked by it error-bars with early-type galaxies being red and dotted, late-type galaxies blue and dashed, and all-type galaxies being black and solid.  The solid black lines represent $c_2 = b_1 - 1.4$, $c_2 = b_1 - 1.6$ and $c_2 = b_1 - 1.8$.  The green open circles display the calculated $c_2$ and $b_1$ using the HOD parameters determined by \cite{Z05}.}
\label{fig:c2b1}
\end{figure}

In Figure \ref{fig:c2b1} we display our calculated and measured $c_2$ and $b_1$, where the open green circles are the calculated values, black solid (all-), red dotted (early-), and blue dashed (late-type) crosses denote all of the measured values, and the three lines represent $c_2 = b_1 - 1.4$,  $c_2 = b_1 - 1.6$, and  $c_2 = b_1 -1.8$.  Using the HOD fits of \cite{Z05} does not predict a strong relationship between $c_2$ and $b_1$.  Changing the redshift weakly affects the calculated bias parameters, as the results show a significant change only when the HOD parameters are changed (the three calculations for $M_r < -19.5$ are hardly distinguishable, as are the three calculations for $M_r < -20.5$).  Comparing these results to each of our measured $c_2$ values for all galaxy types, the calculated values are consistent, though they are all greater than the measured values.  (The agreement is stronger if one considers that changing $\sigma_8$ will shift the values around.)  Both the late-and all-type galaxies appear consistent with $c_2$ being linearly dependent on $b_1$, but the early-type galaxies $c_2$ measurements show no clear dependence on $b_1$.  Our basic modeling suggests that there is nothing unusual about the relationship between $c_2$ and $b_1$ for early-type galaxies.  It further suggests that for each galaxy-type there is redshift evolution of the HOD for a given luminosity.

\subsection{Late-type Galaxies}
\label{sec:LT}
The measurements made for late-type galaxies are significantly different for $z < 0.3$ than for $z > 0.3$ (e.g., Figure \ref{fig:z.3.4A}).  The rise in correlation amplitudes at small scales, first reported by Ross06, happens only for $z > 0.3$.  This suggests that late-type galaxies become much more likely to exist in close groupings at redshifts greater than 0.3.  We are essentially measuring a preponderance of star forming galaxies in tight configurations as the redshift grows larger than 0.3.  This in turn suggests that merger driven star formation becomes common at $z > 0.3$.  This hypothesis is further supported by the fact that the bluest galaxies ($L2$) at $z > 0.3$ display the largest amplitudes at small scales (see Figure \ref{fig:L2}), meaning that the galaxies with the most star formation are most likely to be found in tight groupings.  

This picture is broadly consistent with the concept of downsizing (e.g., \citealt{Cow96}), which essentially states that higher-mass galaxies form stars earlier and more quickly than lower mass galaxies.  The galaxies included in the volume limited samples are relatively bright ($M_r < -19.5$), and thus we find evidence of a preponderance of merger driven star formation only for $z > 0.3$.  Our interpretation is supported by the results of \cite{Heav04} who found that star formation peaked between a redshift of 0.3 and 0.8, using the fossil records of local SDSS galaxies.  Further, they found the star formation rate to be a strong function of galaxy mass, implying that, for $L > L^{*}$, there is little star formation at low redshift  (this is supported by the dearth of late-type galaxies in the volume limited samples at $z < 0.2$ in our current analysis).  This interpretation is further supported by recent results from the Cosmic Evolution Survey \citep{Scoville06}, which found dramatically smaller star formation rates for galaxies $0.20 < z < 0.43$ than for galaxies $0.43 < z < 0.65$.

The implications of $c_{2,late}$ being significantly smaller than that of early-type galaxies can be explained physically (see Tables 1 and 2).  As the over-density of dark matter increases, the over-density of late-type galaxies becomes a smaller percentage of the dark matter over-density.  This is not a surprise, as the well-known morphology-density relationship \citep{Dre80} tells us that the centers of clusters (i.e., the most over-dense regions) are filled with a smaller percentage of late-type galaxies than the outskirts of the clusters.  It naturally follows that late-type galaxies will have a smaller value of $c_{2}$ than early-type galaxies.  In terms of a halo occupation distribution (HOD), one would expect the fraction of late-type galaxies to decrease with the mass of the host dark matter halo, which is the general trend recently determined by \cite{Z05}.  

For $z > 0.3$, $c_{2,late}$ is much closer to $c_{2,early}$ than any of the measurements made at lower redshift (see Tables 1 and 2).  This suggests that at higher redshift, the fraction of red galaxies as a function of density should display a shallower slope than at low redshift.  This is observed by \cite{Yee05}, as for galaxies with $M_r < -19.5$, the slope in this relationship is on average smaller for galaxies with $0.4 < z < 0.6$ than for galaxies with $0.2 < z < 0.4$.  While these redshift ranges are different than the ones we employ, it confirms that the fraction of red galaxies versus density relationship shows a decrease in slope as the redshift increases (for $M_r < -19.5$).  This is in line with the results of \cite{Dress97}, who found that at redshifts $\sim$ 0.5 the fraction of spiral galaxies in clusters is 2-3 times larger than in local clusters and that the spirals at higher redshift essentially replace the S0 fraction.  It is thus likely that the increase in $c_{2,late}$ at $z > 0.3$ is due to cluster spirals that have yet to evolve into S0 galaxies.  At smaller redshifts, the spirals have likely evolved into S0 galaxies.  Galaxies that we classify as late-type are therefore unlikely to be found in dense environments at low redshift, and thus $c_{2,late}$ is significantly smaller for $z < 0.3$ than for $z > 0.3$.

\section{Conclusions}
The results presented in this paper represent the most complete and accurate determination of the $N$-th order correlations of photometrically selected galaxies.  The measurements and the theoretical modeling used to interpret the measurements represent a significant improvement over the Ross06 measurements.  Taking all SDSS galaxies with $18 < r < 21$ and measuring $\bar{\omega}_N$ produces extremely interesting results, but it is only through volume limiting the sample and splitting by type, redshift, and luminosity that we are able to analyze the subtle effects that produce the measurements displayed for all galaxies.  In doing so, we are able to quantify the nature of linear and non-linear clustering, and its dependence on type, redshift, and luminosity.  

We find that the linear bias parameter $b_1$ is smaller for late-type galaxies than for early-type galaxies, a result that is robust against changes in redshift and luminosity, but the ratio of $b_{1,early}$ to $b_{1,late}$ does vary between 1.2 and 1.5 depending on the specific redshift/luminosity bin.  We confirm that $b_1$ increases proportional to luminosity, as found in many previous studies.  Significant evolution appears to occur in galaxies between a redshift of 0.3 and 0.4 as there is a large increase in $b_1$ going between galaxies with $z < 0.3$ and $0.3 < z < 0.4$ but no significant change in $b_1$ between $z =$ 0.2 and $z =$ 0.3.  

The second-order bias, parameterized by $c_2$, is significantly smaller for late- than early-types, and this is robust against any changes in the luminosity, redshift, or $\sigma_8$.  This relationship between the non-linear bias of early- and late-type galaxies can be seen as a rigorous statistical restatement of the density-morphology relationship and agrees with the results of the HOD analysis by \cite{Z05}.  By applying a basic HOD model, we find our measured results are in fair agreement with the HOD parameters determined by \cite{Z05}.  

There are large differences in the correlation measurements of late-type galaxies at redshifts greater- and less-than 0.3.  This is broadly consistent with cosmic downsizing \citep{Cow96}.  These differences predict a great amount of merger driven star formation at $z > 0.3$, and are consistent with the observed evolution in the density/morphology relationship with redshift.  Our results suggest that a detailed study of the density/morphology relationship as a function of redshift would find significant evolution at $z \sim 0.3$.

If we require that bias be linear and set $c_{2,all} = 0$, we find $\sigma_8 = 0.64^{+0.04}_{-0.03}$, consistent with the lower limit on WMAP3 measurements constrained by inflationary models.  If instead, we set $c_{2,all}$ equal to the $\sigma_8$ independent value found by \cite{Gaz05}, we find that $\sigma_8$ is a great match to the WMAP3-SDSS combined measurement.  Considering all of the results, the most likely conclusion is that $c_2$ is at least slightly negative and that $\sigma_8 < 0.8$.

As per usual, the results of our study demand more investigation.  To this end, we are currently working to extend the analyses presented herein in two complementary directions.  First, as was discussed earlier, the value of $c_2$ can be constrained further by correcting the $\bar{\omega}_2$ measurements for the assumed $c_2$ and measured $b_1/\sigma_8$.  Thus, by performing this analysis on the large, homogenous, photometric Luminous Red Galaxy sample from the SDSS and applying the results to our current analysis, we are thereby improving our measurements of higher-order bias terms and their dependence on galaxy-type, redshift, and luminosity. Second, we are also improving our theoretical interpretation of these results by performing a more rigorous halo occupation distribution model analysis of our higher-order correlation function measurements.

\acknowledgements

AJR, RJB and ADM acknowledge support from Microsoft Research, the University of Illinois, and NASA through grants NNG06GH156 and NB 2006-02049.  The authors made extensive use of the storage and computing facilities at the National Center for Supercomputing Applications and thank the technical staff for their assistance in enabling this work.

We thank Ani Thakar and Jan Van den Berg for help with obtaining a copy of the SDSS DR5 databases. 
We thank Enrique Gazta\~{n}aga for helpful discussions of the proper analysis and Tamas Budavari for help describing the {\tt Photoz} table.  We thank Ravi Sheth and Ben Wandelt for helpful comments.

Funding for the creation and distribution of the SDSS Archive has been provided by the Alfred P. Sloan Foundation, the Participating Institutions, the National Aeronautics and Space Administration, the National Science Foundation, the U.S. Department of Energy, the Japanese Monbukagakusho, and the Max Planck Society. The SDSS Web site is http://www.sdss.org/.

The SDSS is managed by the Astrophysical Research Consortium (ARC) for the Participating Institutions. The Participating Institutions are The University of Chicago, Fermilab, the Institute for Advanced Study, the Japan Participation Group, The Johns Hopkins University, the Korean Scientist Group, Los Alamos National Laboratory, the Max-Planck-Institute for Astronomy (MPIA), the Max-Planck-Institute for Astrophysics (MPA), New Mexico State University, University of Pittsburgh, University of Portsmouth, Princeton University, the United States Naval Observatory, and the University of Washington.

\tablewidth{0pt}

\clearpage
\begin{deluxetable}{lccccccc}
\tablecaption{The measured values of bias parameters $b_1$ and $c_2$ for all measurements made using the fifteen $z3$ volume-limited samples.} 
\tablecolumns{8}
\tablehead{ \colhead{Type} & \colhead{$M_r$ Range} & \colhead{z Range} &  \colhead{$h^{-1}$Mpc Range} & \colhead{$b_1$} & \colhead{$\chi^2$/DOF} & \colhead{$c_2$} &   \colhead{$\chi^2$/DOF} 
}
\startdata
All & $<$ -19.5 & $<$ 0.3 & 8 -- 36 & 1.06 $\pm$ 0.01 & 1.10 & -0.45$\pm$ 0.13 & 0.44 \\
All & -19.5 to -20.5 & $<$ 0.3 & 8 -- 36 & 0.97 $\pm$ 0.01 & 0.36 & -0.58$\pm$ 0.20 & 0.29 \\
All & -20.5 to -21.5 & $<$ 0.3 & 8 -- 36 & 1.18 $\pm$ 0.01 & 0.76 & -0.38$\pm$ 0.14 & 0.16 \\
All & $<$ -19.5 & $<$ 0.2 & 8 -- 24 & 1.03 $\pm$ 0.03 & 0.24 & -0.36$\pm$ 0.19 & 0.52 \\
All & $<$ -19.5 & 0.2 to 0.3 & 8.5 -- 38 & 1.03 $\pm$ 0.02 & 1.13 & -0.32$\pm$ 0.15 & 0.29 \\
Early &$<$ -19.5 & $<$ 0.3 & 8 -- 36 & 1.33 $\pm$ 0.03 & 0.31 & 0.06$\pm$ 0.12 & 2.40 \\
Early &-19.5 to -20.5 & $<$ 0.3 & 8 -- 36 & 1.21 $\pm$ 0.02 & 0.51 & 0.15$\pm$ 0.15 & 3.94 \\
Early & -20.5 to -21.5 & $<$ 0.3 & 8 -- 36 & 1.50 $\pm$ 0.03 & 0.17 & 0.08$\pm$ 0.13 & 0.34 \\
Early &$<$ -19.5 & $<$ 0.2 & 8 -- 24 & 1.27 $\pm$ 0.04 & 0.43 & 0.10$\pm$ 0.08 & 1.23 \\
Early &$<$ -19.5 & 0.2 to 0.3 & 8.5 -- 38 & 1.27 $\pm$ 0.03 & 0.18 & 0.05$\pm$ 0.12 & 1.21 \\
Late &$<$ -19.5 & $<$ 0.3 & 8 -- 36 & 0.87 $\pm$ 0.01 & 0.98 & -0.93$\pm$ 0.34 & 0.80 \\
Late &-19.5 to -20.5 & $<$ 0.3 & 8 -- 36 & 0.84 $\pm$ 0.01 & 0.88 & -0.98$\pm$ 0.27 & 0.46 \\
Late & -20.5 to -21.5 & $<$ 0.3 & 8 -- 36 & 1.04 $\pm$ 0.02 & 1.14 & -0.84$\pm$ 0.25 & 0.04 \\
Late &$<$ -19.5 & $<$ 0.2 & 8 -- 24 & 0.79 $\pm$ 0.02 & 0.10 & -1.03$\pm$ 0.35 & 0.27 \\
Late &$<$ -19.5 & 0.2 to 0.3 & 8.5 -- 38 & 0.84 $\pm$ 0.01 & 1.90 & -1.03$\pm$ 0.35 & 0.20 \\

\enddata
  \label{T:bz3}
\end{deluxetable}

\clearpage
\begin{deluxetable}{lccccccc}
\tablecaption{The measured values of bias parameters $b_1$ and $c_2$ for all measurements made using the nine $z4$ volume-limited samples.} 
\tablecolumns{8}
\tablehead{ \colhead{Type} & \colhead{$M_r$ Range} & \colhead{z Range} &  \colhead{$h^{-1}$Mpc Range} & \colhead{$b_1$} & \colhead{$\chi^2$/DOF} & \colhead{$c_2$} &   \colhead{$\chi^2$/DOF} 
}
\startdata
All & $<$ -20.5 & $<$ 0.4 & 10 -- 45 & 1.35 $\pm$ 0.01 & 1.73 & -0.29$\pm$ 0.17 & 0.89 \\
All & $<$ -20.5 & $<$ 0.3 & 8 -- 36 & 1.22 $\pm$ 0.03 & 1.36 & -0.34$\pm$ 0.21 & 0.20 \\
All & $<$ -20.5 & 0.3 to 0.4 & 8 -- 51 & 1.39 $\pm$ 0.01 & 2.81 & -0.17$\pm$ 0.10 & 0.46 \\
Early &$<$ -20.5 & $<$ 0.4 & 10 -- 45 & 1.64 $\pm$ 0.03 & 0.31 & 0.01$\pm$ 0.14 & 1.77 \\
Early &$<$ -20.5 & $<$ 0.3 & 8 -- 36 & 1.53 $\pm$ 0.03 & 0.16 & 0.08$\pm$ 0.12 & 0.70 \\
Early &$<$ -20.5 & 0.3 to 0.4 & 8 -- 51 & 1.66 $\pm$ 0.02 & 1.12 & 0.01$\pm$ 0.08 & 0.88 \\
Late &$<$ -20.5 & $<$ 0.4 & 10 -- 45 & 1.23 $\pm$ 0.02 & 2.20 & -0.52$\pm$ 0.27 & 0.44 \\
Late &$<$ -20.5 & $<$ 0.3 & 8 -- 36 & 1.04 $\pm$ 0.02 & 1.7 & -0.83$\pm$ 0.21 & 0.03 \\
Late &$<$ -20.5 & 0.3 to 0.4 & 8 -- 51 & 1.25 $\pm$ 0.02 & 3.66 & -0.38$\pm$ 0.30 & 0.37 \\

\enddata
  \label{T:bz4}
\end{deluxetable}

\clearpage

\clearpage

\clearpage

\clearpage

\clearpage

\clearpage

\clearpage

\clearpage

\clearpage

\clearpage

\clearpage

\clearpage

\clearpage

\clearpage

\clearpage

\clearpage

\clearpage

\clearpage

\clearpage

\begin{thebibliography}{7}
\bibitem[Abazajian et al. (2003)]{Ab01}Abazajian, K., Adelman, J., Agueros, M., et al.\ 2003, \aj, 126, 2081
\bibitem[Abazajian et al.(2005)]{Ab05} Abazajian, K., et 
al.\ 2005, \aj, 129, 1755
\bibitem[Bernardeau(1994)]{Ber94} Bernardeau, F.\ 1994, \apj, 
433, 1 
\bibitem[Bernardeau et al.(2002)]{Ber02} Bernardeau, F., 
Colombi, S., Gazta{\~n}aga, E., \& Scoccimarro, R.\ 2002, \physrep, 367, 1
\bibitem[Bernardeau(1995)]{Ber95} Bernardeau, F.\ 1995, \aap, 
301, 309
\bibitem[Blaizot et al.(2006)]{Bl06} Blaizot, J., et al.\ 
2006, \mnras, 369, 1009 
\bibitem[Bouchet et al.(1993)]{Bouch93} Bouchet, F.~R., 
Strauss, M.~A., Davis, M., Fisher, K.~B., Yahil, A., \& Huchra, J.~P.\ 
1993, \apj, 417, 36
\bibitem[Budav{\'a}ri et al.(2003)]{Bud03} Budav{\'a}ri, T., 
et al.\ 2003, \apj, 595, 59 
\bibitem[Cole et al.(2005)]{Cole05} Cole, S., et al.\ 2005, 
\mnras, 362, 505
\bibitem[Connolly et al.(2002)]{Con02} Connolly, A.~J., et 
al.\ 2002, \apj, 579, 42 
\bibitem[Cowie et al.(1996)]{Cow96} Cowie, L.~L., Songaila, 
A., Hu, E.~M., \& Cohen, J.~G.\ 1996, \aj, 112, 839 
\bibitem[Croton et al.(2004)]{Cr04} Croton, D.~J., et al.\ 
2004, \mnras, 352, 1232
\bibitem[Croton et al.(2006)]{Cr06} Croton, D.~J., Norberg, 
P., Gaztanaga, E., \& Baugh, C.~M.\ 2006, ArXiv Astrophysics e-prints, 
arXiv:astro-ph/0611313  
\bibitem[Dressler(1980)]{Dre80} Dressler,~A., 1980, \apj, 236, 351
\bibitem[Dressler et al.(1997)]{Dress97} Dressler, A., et al.\ 
1997, \apj, 490, 577 
\bibitem[Eisenstein \& Hu(1998)]{Eis98} Eisenstein, D.~J., \& 
Hu, W.\ 1998, \apj, 496, 605 
\bibitem[Frith et al.(2005a)]{FR05b} Frith, W.~J., Outram, 
P.~J., \& Shanks, T.\ 2005a, \mnras, 364, 593 
\bibitem[Frith, Outram, \& Shanks (2005b)]{FR05} Frith, W.~J., Outram,~P.~J., \& Shanks,~T. \ 2005b, \mnras, accepted, {\it astro-ph/0507704}
\bibitem[Fry \& Gaztanaga(1993)]{Fry93} Fry, J.~N., \& 
Gaztanaga, E.\ 1993, \apj, 413, 447 \bibitem[Fukugita et al.(1996)]{F} Fukugita, M., 
Ichikawa, T., Gunn, J.~E., Doi, M., Shimasaku, K., \& Schneider, D.~P.\ 
1996, \aj, 111, 1748
\bibitem[Gaztanaga(1992)]{Gaz92} Gaztanaga, E.\ 1992, \apjl, 
398, L17 
\bibitem[Gazta\~{n}aga(1994)]{Gaz94} Gazta\~{n}aga, E.\ 1994, \mnras, 
268, 913 
\bibitem[Gaztanaga \& Bernardeau(1998)]{Gaz98} Gaztanaga, E., 
\& Bernardeau, F.\ 1998, \aap, 331, 829 
\bibitem[Gazta{\~n}aga(2002a)]{GazEDRa} Gazta{\~n}aga, E.\ 2002, 
\apj, 580, 144 
\bibitem[Gazta{\~n}aga(2002b)]{GazEDR} Gazta{\~n}aga, E.\ 2002, 
\mnras, 333, L21 
\bibitem[Gazta{\~n}aga et al.(2005)]{Gaz05} Gazta{\~n}aga, 
E., Norberg, P., Baugh, C.~M., \& Croton, D.~J.\ 2005, \mnras, 364, 620 
\bibitem[Goto et al.(2003)]{Go03} Goto,~T., Yamauchi,~C., Fujita,~Y., Okamura,~S., Sekiguchi,~M., Smail,~I., Bernardi,~M., \& Gomez,~P.~L., 2003, \mnras, 346, 601
\bibitem[Groth \& Peebles(1977)]{Gro77} Groth, E.~J., \& 
Peebles, P.~J.~E.\ 1977, \apj, 217, 385 
\bibitem[Gunn et al.(1998)]{C} Gunn, J.~E., et al.\ 1998, 
\aj, 116, 3040
\bibitem[Hamilton A. J. S. (1992)]{H92} Hamilton A. J. S.,\ 1992, \apj, 385, L5
\bibitem[Heavens et al.(2004)]{Heav04} Heavens, A., Panter, 
B., Jimenez, R., \& Dunlop, J.\ 2004, \nat, 428, 625 
\bibitem[Hikage et al.(2005)]{H05} Hikage, C., Matsubara, 
T., Suto, Y., Park, C., Szalay, A.~S., \& Brinkmann, J.\ 2005, \pasj, 57, 
709 
\bibitem[Juszkiewicz et al.(1993)]{J93} Juszkiewicz, R., 
Bouchet, F.~R., \& Colombi, S.\ 1993, \apjl, 412, L9
\bibitem[Kaiser N. (1987)]{K87} Kaiser N., \ 1987, \mnras, 227, 1
\bibitem[Lupton et al.(2002)]{L} Lupton, R.~H., Ivezic, 
Z., Gunn, J.~E., Knapp, G., Strauss, M.~A., \& Yasuda, N.\ 2002, \procspie, 
4836, 350
\bibitem[Madgwick et al.(2003)]{Ma03} Madgwick, D.~S., et 
al.\ 2003, \mnras, 344, 847 
\bibitem[Myers et al.(2003)]{Mye03} Myers,~A.~D., Outram,~P.~J., Shanks,~T., Boyle,~B.~J., Croom,~S.~M., Loaring,~N.~S., Miller,~L., \& Smith,~R.~J. 2003, \mnras, 342, 467
\bibitem[Myers et al.(2005)]{Mye05} Myers,~A.~D., Outram,~P.~J., Shanks,~T., Boyle,~B.~J., Croom,~S.~M., Loaring,~N.~S., Miller,~L., \& Smith,~R.~J. 2005, \mnras, 359, 741 
\bibitem[Myers et al.(2006)]{Mye06} Myers, A.~D., et al.\ 
2006, \apj, 638, 622
\bibitem[Myers et al.(2007)]{Mye07} Myers, A.~D., et al.\ 
2007, \apj, 658, 85 
\bibitem[Nichol et al.(2006)]{Nic06} Nichol, R.~C., et al.\ 
2006, \mnras, 368
\bibitem[Nishimichi et al.(2006)]{Nish06} Nishimichi, T., 
Kayo, I., Hikage, C., Yahata, K., Taruya, A., Jing, Y.~P., Sheth, R.~K., \& 
Suto, Y.\ 2006, ArXiv Astrophysics e-prints, arXiv:astro-ph/0609740 
\bibitem[Norberg et al.(2002)]{N02} Norberg, P., et al.\ 
2002, \mnras, 332, 827
\bibitem[Pan \& Szapudi(2005)]{Pan05} Pan, J., \& Szapudi, 
I.\ 2005, \mnras, 362, 1363
\bibitem[Peebles(1980)]{P80} Peebles, P.~J.~E.\ 1980, 
Research supported by the National Science Foundation.~Princeton, N.J., 
Princeton University Press, 1980.~435 p.
\bibitem[Percival et al.(2001)]{Per01} Percival, W.~J., et 
al.\ 2001, \mnras, 327, 1297 
\bibitem[Rojas et al.(2004)]{Roj04} Rojas, R.~R., Vogeley,~M.~S., Hoyle,~F., \& Brinkmann,~J., 2004, \apj, 617, 50
\bibitem[Riess et al.(2004)]{Riess04} Riess, A.~G., et al.\ 
2004, \apj, 607, 665 
\bibitem[Ross et al.(2006)]{Ross} Ross, A.~J., Brunner, 
R.~J., \& Myers, A.~D.\ 2006, \apj, 649, 48
\bibitem[Saunders et al.(1991)]{Saund91} Saunders, W., Frenk, 
C., Rowan-Robinson, M., Lawrence, A., \& Efstathiou, G.\ 1991, \nat, 349, 
32 
\bibitem[Schlegel, Finkbeiner \& Davis(1998)]{Sc} Schlegel, D.~J., 
Finkbeiner, D.~P., \& Davis, M.\ 1998, \apj, 500, 525
\bibitem[Scoville et al. (2006)]{Scoville06} Scoville, N., et al.\ 2006, ArXiv Astrophysics e-prints, arXiv:astro-ph/0612384
\bibitem[Scranton et al.(2002)]{Scr02} Scranton, R., et al.
2002, \apj, 579, 48 
\bibitem[Seljak et al.(2005)]{Sel05} Seljak, U., et al.\ 
2005, \prd, 71, 043511
\bibitem[Sheth \& Tormen(1999)]{Sheth99} Sheth, R.~K., \& 
Tormen, G.\ 1999, \mnras, 308, 119
\bibitem[Sheth et al.(2001)]{Sheth01} Sheth, R.~K., Mo, H.~J., 
\& Tormen, G.\ 2001, \mnras, 323, 1  
\bibitem[Smith et al.(2003)]{Smith} Smith, R.~E., et al.\ 
2003, \mnras, 341, 1311  
\bibitem[Spergel et al.(2003)]{Sper03} Spergel, D.~N., et al.\ 
2003, \apjs, 148, 175
\bibitem[Spergel et al.(2006)]{Sper06} Spergel, D.~N., et al.\ 
2006, ArXiv Astrophysics e-prints, arXiv:astro-ph/0603449 

\bibitem[Stoughton et al.(2002)]{Sto02} Stoughton,~C.,~et~al. 2002, \aj, 123, 485
\bibitem[Strateva et al.(2001)]{Stra01} Strateva, I., et al.\ 
2001, \aj, 122, 1861 
\bibitem[Szapudi et al.(1992)]{Sza92} Szapudi, I., Szalay, 
A.~S., \& Boschan, P.\ 1992, \apj, 390, 350 
\bibitem[Szapudi et al.(2002)]{Sza02} Szapudi, I., et al. 2002, \apj, 570, 75 
\bibitem[Tegmark et al.(2002)]{Teg02} Tegmark, M., et al.\ 
2002, \apj, 571, 191
\bibitem[Tegmark et al.(2004)]{Teg04} Tegmark, M., et al.\ 
2004, \prd, 69, 103501 
\bibitem[Voevodkin \& Vikhlinin(2004)]{Vod04} Voevodkin, A., 
\& Vikhlinin, A.\ 2004, \apj, 601, 610 
\bibitem[Willmer et al.(1998)]{W98} Willmer, C.~N.~A., da 
Costa, L.~N., \& Pellegrini, P.~S.\ 1998, \aj, 115, 869
\bibitem[York et al.(2000)]{Y} York, D.~G., et al. 2000, 
\aj, 120, 1579 
\bibitem[Yee et al.(2005)]{Yee05} Yee, H.~K.~C., Hsieh, 
B.~C., Lin, H., \& Gladders, M.~D.\ 2005, \apjl, 629, L77 
\bibitem[Zehavi et al.(2002)]{Z02} Zehavi, I., et al.\ 
2002, \apj, 571, 172
\bibitem[Zehavi et al.(2005)]{Z05} Zehavi, I., et al.\ 
2005, \apj, 630, 1 
\end{thebibliography}
\end{document}